\documentclass[reprint,aps,prb,epsf,superscriptaddress,amsmath,amssymb,amsfonts,showpacs]{revtex4-2}

\include{generalSettings}
\usepackage[utf8]{inputenc}
\usepackage{graphicx}
\usepackage{epsfig}
\usepackage{dcolumn}
\usepackage{bm}
\usepackage{braket}
\usepackage{amsmath}
\usepackage{titlesec}
\usepackage{placeins}
\usepackage{color}
\usepackage{subfig}
\graphicspath{{./}{./Figures_latex/}}

\newcommand{\beginsupplement}{%
	\setcounter{table}{0}
	\renewcommand{\thetable}{A\arabic{table}}%
	\setcounter{figure}{0}
	\renewcommand{\thefigure}{A\arabic{figure}}%
	\setcounter{equation}{0}
	\renewcommand{\theequation}{A\arabic{equation}}%
}

\begin{document}
	
\title{Creating arbitrary sequences of mobile magnetic skyrmions and antiskyrmions}
\author{Pia Siegl}
\affiliation{I. Institut für Theoretische Physik,  Universität Hamburg, Notkestr.\ 9, 22607 Hamburg, Germany}%
\affiliation{The Hamburg Centre for Ultrafast Imaging,  Universität Hamburg, Luruper Chaussee 149, 22761 Hamburg, Germany}

\author{Martin Stier}
\affiliation{I. Institut für Theoretische Physik,  Universität Hamburg, Notkestr.\ 9, 22607 Hamburg, Germany}%
\affiliation{The Hamburg Centre for Ultrafast Imaging,  Universität Hamburg, Luruper Chaussee 149, 22761 Hamburg, Germany}

\author{Alexander F. Schäffer}
\affiliation{I. Institut für Theoretische Physik,  Universität Hamburg, Notkestr.\ 9, 22607 Hamburg, Germany}%
\affiliation{Fachbereich Physik,  Universität Hamburg, Jungiusstr.\ 11a, 20355 Hamburg, Germany}%
\affiliation{Institute of Physics, Martin-Luther-Universität Halle-Wittenberg, 06120, Halle (Saale), Germany}

\author{Elena Y. Vedmedenko}
\affiliation{The Hamburg Centre for Ultrafast Imaging,  Universität Hamburg, Luruper Chaussee 149, 22761 Hamburg, Germany}
\affiliation{Fachbereich Physik,  Universität Hamburg, Jungiusstr.\ 11a, 20355 Hamburg, Germany}%

\author{Thore Posske}
\affiliation{I. Institut für Theoretische Physik,  Universität Hamburg, Notkestr.\ 9, 22607 Hamburg, Germany}%
\affiliation{The Hamburg Centre for Ultrafast Imaging,  Universität Hamburg, Luruper Chaussee 149, 22761 Hamburg, Germany}

\author{Roland Wiesendanger}
\affiliation{The Hamburg Centre for Ultrafast Imaging,  Universität Hamburg, Luruper Chaussee 149, 22761 Hamburg, Germany}
\affiliation{Fachbereich Physik,  Universität Hamburg, Jungiusstr.\ 11a, 20355 Hamburg, Germany}%

\author{Michael Thorwart}
\affiliation{I. Institut für Theoretische Physik,  Universität Hamburg, Notkestr.\ 9, 22607 Hamburg, Germany}%
\affiliation{The Hamburg Centre for Ultrafast Imaging,  Universität Hamburg, Luruper Chaussee 149, 22761 Hamburg, Germany}

\begin{abstract}
	Magnetic skyrmions and their anti-particles, the antiskyrmions, are stable magnetic solitons existing down to the nanometer scale. Their stability and size as well as the possibility to propel them by, e.g., electric currents make them promising candidates for the use in memory devices, such as racetrack memories. Skyrmions and antiskyrmions share those same advantages individually, but may annihilate each other when they coexist in the same device. Yet, combining them to represent logical bits of 0 (skyrmion) and 1 (antiskyrmion) in one device opens new possibilities to create new types of densely packed racetrack memory devices.
	For this, a controlled creation and annihilation procedure, i.e., a writing or deletion operation, is necessary. Here, we propose a method to create arbitrary sequences of coexisting skyrmions and antiskyrmions by rotations of the magnetic moments at the edge of a rectangular slab. The skyrmions and antiskyrmions remain stable and do not annihilate each other.
\end{abstract}

\maketitle

\section{Introduction}
Magnetic skyrmions (SKs) are promising candidates for reliable, energy efficient, and high density data storage \cite{roadmap} in the form of racetrack memory elements \cite{fert2013skyrmions,muller2017magnetic}. Skyrmions and their anti-particles, the antiskyrmions (ASKs) \cite{kovalev2018skyrmions,nayak2017magnetic,jena2019observation}, are magnetic solitons which may appear on a nanometer scale while remaining surprisingly stable \cite{everschor2018perspective}. This stability is typically well explained by topological arguments \cite{Nagaosa2013}, even though the necessary continuous transformations are not possible on a physical atomic lattice \cite{Vedmedenko2019}. In particular, the phenomena of SK creation or annihilation are beyond the scope of the static concept of topology. Thus, these processes have to be discussed theoretically by energy considerations \cite{PhysRevB.94.174418} or computational simulations \cite{lin2013manipulation}.
In recent years, several creation mechanisms have been found or predicted, such as reversible creation/annihilation by local spin current injection \cite{romming2013writing}, electric fields \cite{Hsu2017, electricfieldinduced} or edge manipulation \cite{Schaffer2020}. It is also possible to create SKs by in-plane currents \cite{stier2017skyrmion,everschor2017skyrmion}, out-of-plane currents \cite{Yuan2016}, spin waves \cite{spinwavecreation2015}, lasers \cite{laserinducedcreation,laserinducedcreation2,koshibae2014creation} or geometric constraints \cite{jiang2015blowing}.\\
The creation of SKs obviously lays the foundation for SK based devices, but a precise control over the SK dynamics remains as important. Particularly for the use of racetrack devices, SKs are typically moved by electrical currents \cite{iwasaki2013current} while also magnetic fields \cite{Zhang2018,moon2016skyrmion} or a edge-induced pushing \cite{Schaffer2020} may be used. Notable obstacles for precise SK dynamics in terms of technical usability are the topological SK Hall effect \cite{chen2017skyrmion,jiang2017direct,plettenberg2020steering} or material impurities \cite{potkina2020skyrmions,stier2021role}. It is also important to spatially separate the logical bits within a memory device which lead to the proposal of two-lane racetracks \cite{muller2017magnetic, song2017skyrmion,plettenberg2020steering} or the simultaneous use of two different data carriers as domain walls and SKs \cite{Schaffer2020}. Antiskyrmions would be natural candidates to complement SKs in a two-species racetrack as they share the same advantages, such as, stability, mobility and small size. This, however, imposes the question how to create ASKs in a controlled manner.\\
Theoretically, an ASK is not fundamentally different from a SK in terms of creation. Actually, a creation process often involves meta-stable SK-ASK pairs, where one partner decays due to dissipation and the other remains \cite{stier2017skyrmion,everschor2017skyrmion}. The creation of coexisting stable SKs and ASKs, however, is difficult because materials including Dzayloshinskii-Moriya interactions (DMI) are mostly known to host either SKs or ASKs, depending on the DMI \cite{Hoffmann2017}. Energetic equivalence is only predicted for one-dimensional DMI \cite{Hoffmann2017}. Coexisting lattices of SKs and ASKs were reported for a small temperature pocket in Heusler compounds \cite{Jena2020}.
Furthermore, the creation of both SKs and ASKs by local heating in dipolar magnets has been numerically simulated \cite{koshibae2014creation}.
However, both of these processes rely on the stochastic formation of SKs and ASKs, not bearing any control of which species is created. By anisotropy design, a controlled creation of SKs and ASKs can be achieved at fixed positions \cite{zhang2016creation}, preventing, however, any mobility of the data carriers, which may be an important disadvantage for the design of memory devices.\\
In this work, we propose a method for the controlled creation of coexisting mobile SKs and ASKs. This allows for the writing and deletion of arbitrary sequences of densely packed SKs and ASKs, which may be efficiently used in racetrack devices. We consider magnetic systems with anisotropic DMI and numerically simulate their dynamics by solving the Landau-Lifshitz-Gilbert equation. Hereby, we show that both magnetic species can be created by a rotation of the edge magnetization, where the sense of rotation determines which quasiparticle species is formed. Finally, we discuss parameter regimes where this rotation scheme is feasible.\\
This work is organized as follows.
In Sec. \ref{SecModel}, we introduce our model, a magnetic lattice with nearest neighbor interactions, and the Landau-Lifshitz-Gilbert equation used to compute the time evolution. 
In Sec. \ref{SecCreationSingleSk}, we propose a creation scheme for SKs and ASKs in the same material with a given one-dimensional DMI, relying on a rotation of the edge magnetization.
Further, we extend the discussion in Sec. \ref{SecSkArray} to the creation of an array of coexisting mobile SKs and ASKs. In Sec. \ref{Sec2DDMI}, we show that the presented schemes work beyond a one-dimensional DMI.
Finally, we present our conclusion and outlook in Sec. \ref{SecConclusion}.

\section{Model}\label{SecModel}
We consider an $N_x\times N_y$ magnetic square lattice of normalized magnetic moments $\textbf{n}_{\textbf r}(t)$ with ferromagnetic boundaries $\textbf{n}_{\textbf r\in \text{edge}}(t)= -1$.
All interactions are collected in the atomistic lattice Hamiltonian
\begin{align}\label{Lattice_Hamiltonian}
H=&-J\sum_{\textbf{r}}\textbf{n}_{\textbf{r}}\cdot(\textbf{n}_{\textbf{r}+a\hat{x}}+\textbf{n}_{\textbf{r}+a\hat{y}})\nonumber\\
&-B_z\sum_{\textbf{r}}n^z_{\textbf{r}}
-K\sum_{\textbf{r}}(n^z_{\textbf{r}})^2\\
&-\sum_{\textbf{r}}\left[D_x(\textbf{n}_{\textbf{r}} \times \textbf{n}_{\textbf{r}+a\hat{y}})\cdot\hat{x}-D_y(\textbf{n}_{\textbf{r}} \times \textbf{n}_{\textbf{r}+a\hat{x}})\cdot\hat{y}\right]\nonumber .
\end{align}
To be specific, we consider interaction parameters of PdFeIr(111) adapted to a square lattice \cite{Romming2015,Schaffer2020} and with an adapted DMI. That corresponds to the exchange interaction strength $J=11.6$~meV, the DMI strengths $|D_x|\leq D_y$ and $D_y=3.17$~meV, the magnetic anisotropy $K=0.35$~meV and the external magnetic field strength $B_z=-0.261$~meV.\\
We compute the time evolution of the system by solving the extended Landau-Lifshitz-Gilbert equation \cite{Tatara2008,Bazaliy1998,Zhang2004,Lakshmanan2011} 
\begin{equation}\label{LLG_current}
\begin{split}
\frac{\partial \textbf{n}}{\partial t}=&-\frac{\gamma}{1+\alpha^2}~\textbf{n}\times\textbf{B}_\textbf{eff}-\frac{\alpha \gamma}{1+\alpha^2}~ \textbf{n}\times ( \textbf{n}\times\textbf{B}_\textbf{eff})\\
&+(1+\alpha \beta)(\textbf{v}_s\cdot\nabla)\textbf{n}
+(\alpha-\beta)\textbf{n}\times(\textbf{v}_s\cdot\nabla)\textbf{n},
\end{split}
\end{equation}
with a fourth order Runge-Kutta method where $\alpha$ is the damping constant, $\beta$ the non-adiabaticity constant, $\gamma$ the gyromagnetic ratio and $\textbf{v}_s=\frac{pa^3}{2e}\textbf{j}$ the effective spin velocity with the polarization $p$, the lattice constant $a$, the elementary charge $e>0$ and the in-plane electric charge current density $\textbf{j}$.
The effective field is $\textbf{B}_{\text{eff}}=-\frac{1}{\gamma\hbar}\frac{\partial H}{\partial\textbf{n}}$.\\
\section{Creation process of single (anti)skyrmions}\label{SecCreationSingleSk}
\begin{figure*}[!tbp]
	\centering
	\includegraphics[width=\linewidth, height=\textheight,keepaspectratio]{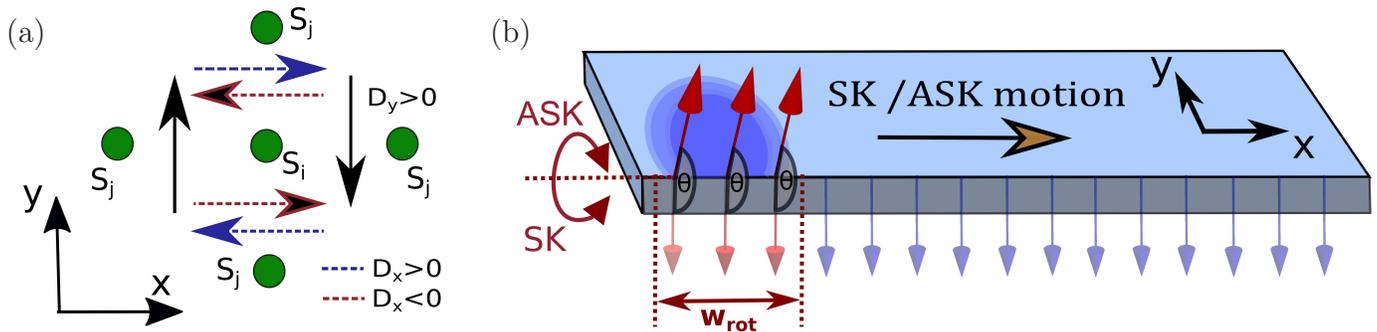}
	\begin{picture}(0,0)
	\put(-260,120){{\large (a)}}
	\end{picture}
	\begin{picture}(0,0)
	\put(-80,120){{\large (b)}}
	\end{picture}
	\caption{(a) Orientation of the DMI.  For a fixed positive $D_y$, a positive $D_x$ in Eq. (\ref{Lattice_Hamiltonian}) stabilizes a skyrmion (blue arrows), a negative $D_x$ an antiskyrmion (red arrows). We note that $D_x$ induces a twist in the lateral $y$ direction, as $D_y$ does in the lateral $x$ direction. (b) Schematic of a SK creation by a rotation of a stripe of the edge magnetization (red arrows). Except for the rotated stripe of magnetization, the sample has a boundary with parallel spin configuration (blue arrows). Depending on the sense of rotation a SK or ASK is created and afterwards pushed along the racetrack.}
	\label{Fig1}
\end{figure*}

\begin{figure*}[!tbp]\centering
	\includegraphics[width=\linewidth, height=\textheight,keepaspectratio]{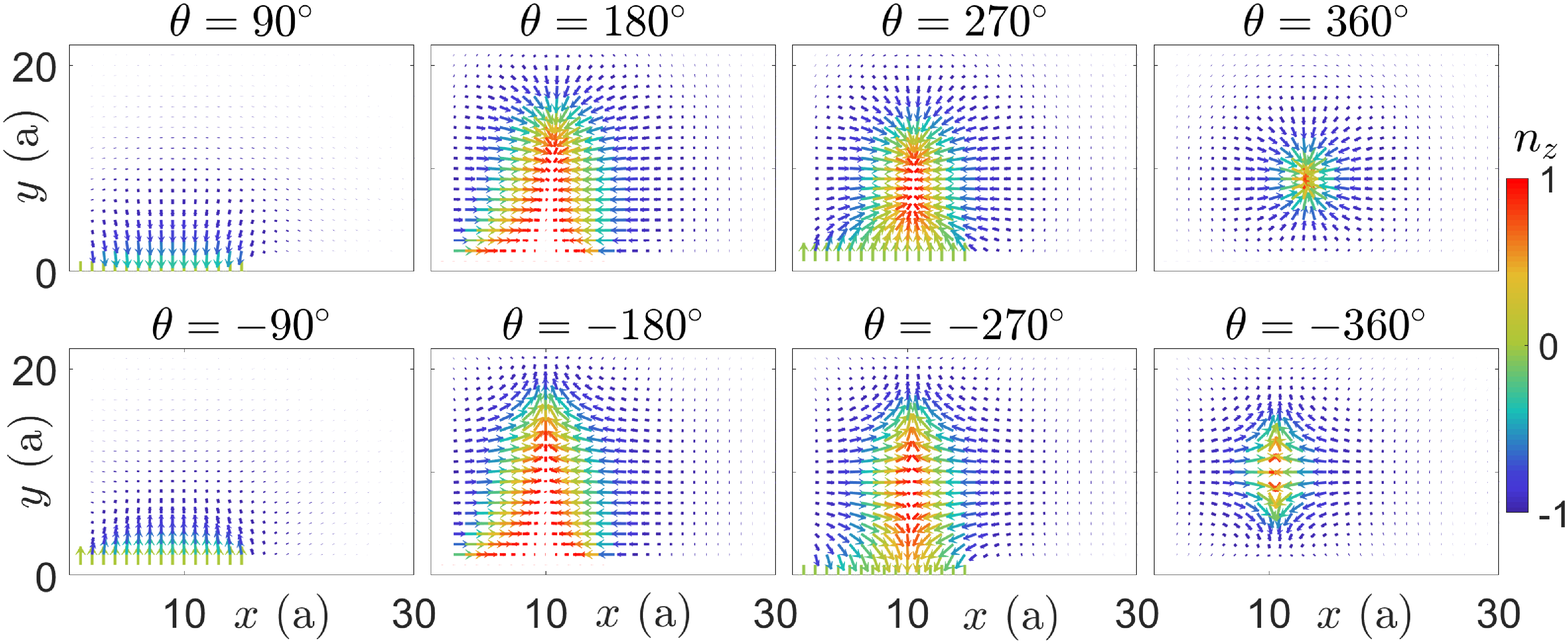}
	\caption{ Creation process of a skyrmion (top row) and an antiskyrmion (bottom row) at different rotation angles $\theta=\pm\nu t$. The different signs of $\theta$ reflect the different senses of rotation. The arrows depict the $x-y$  orientation of the magnetization over the position, the $z$-component is included via the color code. The edge magnetization is rotated with $\nu=2.78$~GHz while $D_x =0$.}
	\label{Fig2}
\end{figure*}

Ref. \cite{Schaffer2020} shows, that in systems with isotropic DMI, i.e., $D_x=D_y$, it is possible   to create coexisting SKs and magnetic domain walls by a rotation of magnetic moments at the edge. However, a racetrack design based solely on skyrmionic quasiparticles promises  smaller, more mobile data carriers that are and less prone to material impurities. From Eq. (\ref{Lattice_Hamiltonian}) we find, that the DMI prefers twisted magnetic structures with the part pertaining to $ D_x$ favoring a rotation in the lateral $y$ direction and the one pertaining to $ D_y$ in lateral $x$ direction. Since the twist of the magnetization of a SK and an ASK coincides only in one direction, they are usually stabilized by different orientations of the DMI, as shown in Fig. \ref{Fig1} (a). A DMI with $D_x/D_y>0$ favors a SK and a DMI with $D_x/D_y<0$ an ASK \cite{Huang2017}. Thus, SKs and ASKs are typically mutually exclusive. Only in the special case $D_x=0$ or $D_y=0$, both species are energetically equivalent \cite{Hoffmann2017}.\\
For the actual creation process, we focus first on $D_x=0$ to have a strictly one-dimensional DMI. Then, the twist of the magnetization  of the (anti)skyrmion is only prescribed in $x$ direction by the DMI as $D_y\neq 0$, while it is not determined in $y$ direction by the DMI. This opens possibility to imprint a twist by the edge rotation. As discussed in Ref. \cite{Schaffer2020}, there is a manifold of boundary rotations that can create SKs. Thus, in real materials, the explicit rotation can be chosen as technically suitable and can be performed by, e.g., repeated current pulses as in toggle magnetoresistive random access memory or a coupling to twisted magnetic microscopic structures. For an extensive discussion of possible implementations of the rotation, we refer to a previous work \cite{Schaffer2020}. Here, we use a uniform rotation for simplicity which is given by
\begin{equation}\label{RotEq}
\begin{split}
&n^x_{\text{rot}}(t) = 0,\\
&n^y_{\text{rot}}(t) = \pm
\sin(\nu t),\\
&n^z_{\text{rot}}(t) = -\cos(\nu t),
\end{split}
\end{equation} 
on a stripe of length $w_{\text{rot}}=15a$ at the start of the racetrack geometry, as indicated in Fig. \ref{Fig1}.
Depending on the sense of rotation of $n^y_{\text{rot}}(t)$, a SK or an ASK is created. The existence and speed of the specific creation process obviously depends on several model parameters, where a detailed analysis of these dependencies is beyond the scope of this work. A more detailed discussion about possible rotation frequencies can be found in Ref. \cite{Schaffer2020}. Here, an exemplary creation process for both species is shown in Fig. \ref{Fig2} for $\nu=2.78$ GHz which yields a creation in less than a nanosecond. As expected from the results in Ref. \cite{Schaffer2020}, we also see in our current simulations that the created species can be deleted by rotating again in backwards direction if the racetrack width $N_y$ is not too large. For large $N_y$, a SK-ASK hybrid can be created during the intended deletion process due to the large extension in $y$-direction of the quasiparticle. If the species is not present, the respective antiparticle is created by the backwards rotation. In Fig.\ \ref{Fig2}, we have confined the wire in $y$ direction as it prevents an elongation of the (anti)skyrmions which would occur due to $|D_x|<D_y$.\\
\section{Creation process of a skyrmion/antiskyrmion array}\label{SecSkArray} 
\begin{figure}[!tb]\centering
	\includegraphics[width=\linewidth, height=\textheight,keepaspectratio]{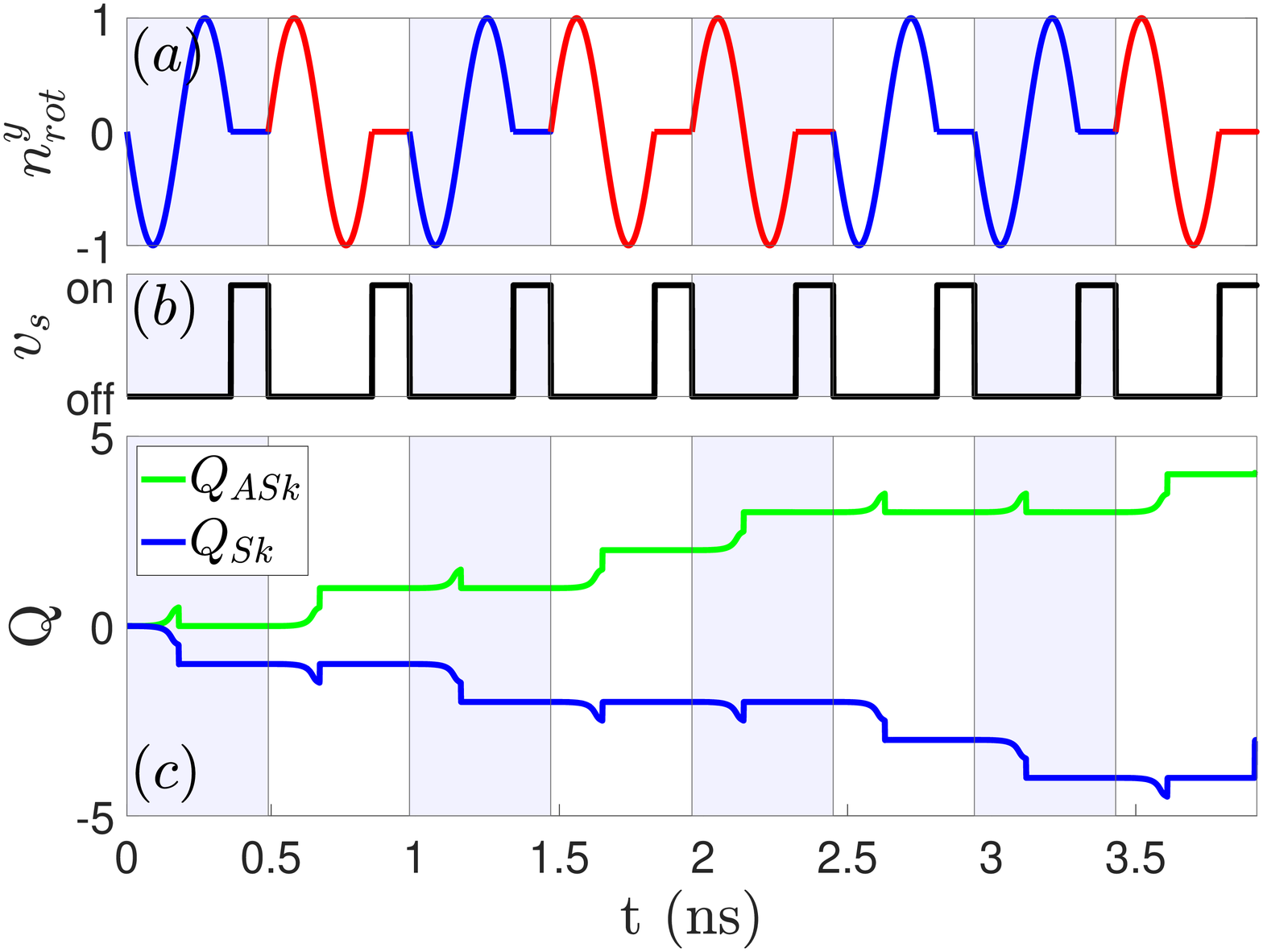}
	\includegraphics[width=\linewidth, height=\textheight,keepaspectratio, trim=40 150 0 180,clip]{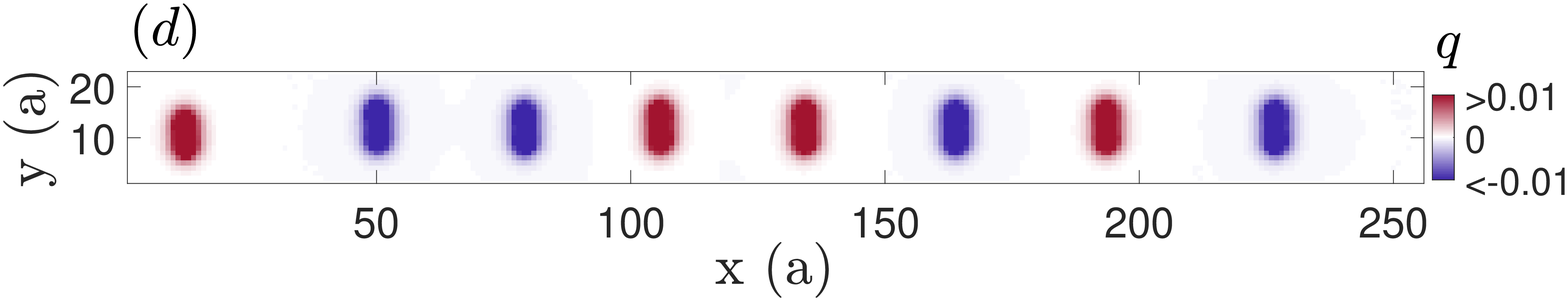}
	\caption{Creation protocol of a skyrmion-antiskyrmion array of eight quasiparticles.
		(a) The time evolution of the $y$-component of the rotated edge magnetization $n_{rot}^y$ (Eq. (\ref{RotEq})) reflects the creation processes of SKs (blue) and ASKs (red).  (b)  Between two creation events a current pulse is applied to move the created quasiparticles along the sample. (c) Evolution of the negative and positive topological charge contributions $Q_{Sk}$ and $Q_{ASk}$, which count the number of SKs and ASKs, respectively. Here, each change of the topological charge corresponds to a creation process. After each creation process, a plateau is reached at an integer value, indicating a stable number of SKs and ASKs. (d) The spatial distribution of the topological charge density at $t=3.8$~ns depicts eight stable SKs and ASKs.}
	\label{Fig3}
\end{figure}

To store information with the help of SKs and ASKs in a racetrack, a controlled creation, deletion and propulsion of both magnetic species in an arbitrary order is necessary.
In a pure SK system, an array of SKs can be created by repeating the same rotation process \cite{Schaffer2020}. Repulsive interactions between the SKs shift the already created SKs along the wire during each new creation process.
However, the interaction between SKs and ASKs depends on the orientation towards each other and can be either repulsive or attractive \cite{Hoffmann2021, Shimizu2020}. An attractive interaction usually leads to an annihilation of both species.
In the setup presented here, SKs and ASKs repel each other in $x$ direction while they would annihilate in $y$ direction. As they are, in contrast to Ref. \cite{Schaffer2020}, created at the long edge along the $x$ axis, successively created SKs and ASKs would be positioned along the $y$ axis and annihilate each other.
Thus, to successively create and stabilize SKs and ASKs, the already created quasiparticles need to be moved along the sample ($x$ direction) before the next creation process takes place.
Here, we achieve this by applying a current along the $x$ direction while currentless racetrack realizations are also possible (see Appendix).
To track the creation of SKs and ASKs, we compute the topological charge 
\begin{equation}
Q=\frac{1}{4\pi}\iint\underbrace{ \textbf{n}\cdot\left(\frac{\partial\textbf{n}}{\partial x}\times\frac{\partial\textbf{n}}{\partial y}\right)}_{= q(x,y,t)}\text{d}x\text{d}y,
\end{equation}
where $q(x,y,t)$ is the topological charge density. Note that the topological charge has to take integer values for ferromagnetic boundary conditions, but is not restricted to this during the rotation where the boundary conditions change. As SKs and ASKs have a topological charge of opposite sign, the total topological charge does only provide information about the difference in the number of SKs and ASKs. We therefore also consider $Q_{Sk}$ ($Q_{ASk}$) which is the sum over the negative (positive) contributions of the topological charge density and thus approximates the number of SKs (ASKs). Again, $Q_{Sk}$ or $Q_{ASk}$ do not have to take integer values as topological arguments only hold for the total topological charge $Q$. However, since the DMI leads to a repulsive interaction along the $x-$ axis, the SKs and ASKs remain clearly separated which yields values for $Q_{Sk}$ or $Q_{ASk}$ reasonably close to integers when the boundary magnetization is not tilted\\
The creation protocol of eight SKs and ASKs is shown in Fig. \ref{Fig3}. Between each creation process a rotation break of duration $\tau=130$~ps is taken. During this break, a current of $v_s=-0.05$~a/ps is applied in $x$ direction. For simplicity, we consider $\beta=\alpha=0.1$, such that no SK Hall effect appears for different values of $\beta$, see Appendix.
After completing eight rotations, an array of in total eight SKs and ASKs is stabilized in the sample, as shown in Fig. \ref{Fig3}(d).
As mentioned above, a quasiparticle can also be deleted at the edge with a rotated magnetization by reversing the sense of rotation if the racetrack width $N_y$ is not too large. The presented setup therefore allows for a controlled creation, deletion and propulsion of SK-ASK arrays of arbitrary order. Additionally, this racetrack design does not rely on voids between SKs as a logical ``0'' but on two different quasiparticles. Thus, it does not suffer from the necessity to keep two particles in a certain distance as conventional one-lane racetracks relying on, e.g., SKs as information carriers only \cite{Fert2013}. 
The readout of SKs and ASKs is expected to be possible with the help of the topological Hall effect \cite{kovalev2018skyrmions, Kumar2020} or the noncollinear Hall effect \cite{Bouaziz2021}.
\\ 
\section{Two-dimensional DMI}\label{Sec2DDMI}
\begin{figure}[!tb]\centering
	\includegraphics[width=\linewidth, height=\textheight,keepaspectratio]{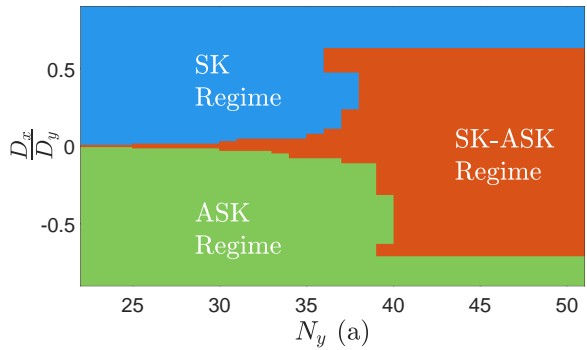}
	\caption{\label{DxDy-Ny}Stability diagram of a skyrmion-antiskyrmion array of five quasiparticles for different values of $\frac{D_x}{D_y}$ and racetrack widths $N_y$ for $\nu=2.78$~GHz, $\tau=150$~ps and $v_s=-0.05$~a/ps. In the ``SK-ASK Regime'' an array of five quasiparticles can be stably inscribed and moved along the racetrack. In the ``SK Regime'' (``ASK Regime'') ASKs (SKs) decay during the creation process or the propulsion.}
\end{figure}

For a one-dimensional DMI, SKs and ASKs are energetically equivalent \cite{Hoffmann2017}. 
As soon as $D_x\neq0$, one of the two quasiparticles becomes energetically favored. 
Even though the DMI is subject to a certain tunability \cite{Tacchi2017,Cho2015,Camosi2017}, the realization of a strictly one-dimensional DMI would pose a hard challenge. Instead, there are various materials that host an anisotropic DMI with significant differences in the magnitude of the components. One of them is Au/Co/W(100), where Camosi {\em et al.\/} \cite{Camosi2017} could measure a DMI ratio of $1:3$.
Next, we show that the controlled creation of SKs and ASKs can be achieved for $D_x\neq 0$ as well.
Figure \ref{DxDy-Ny} depicts the regime where a SK-ASK array of five quasiparticles remains stable for different contributions of $D_x$ and different sample widths, where actually three of the five quasiparticles are of the energetically unfavored type.
We find that for a sufficiently large sample width of $N_y\geq 40$, an array of SKs and ASKs can be stably created and moved, up to a DMI ratio of  $\left|\frac{D_x}{D_y}\right|=0.6$. The stability regime is studied by considering the final number of magnetic structures, when the simulation is performed long enough that the array reaches an equilibrium position after the creation and propulsion.
We confirmed by additional calculations that the SK-ASK regime remains the same for longer arrays. 
For smaller values of $N_y$ or larger ratios of $|D_x/D_y|$, only one magnetic species can be stably inscribed. When $D_x$ and $D_y$ share the same sign, a SK regime appears, where SKs can be created in a controlled manner, while ASKs decay either within the creation process or during the current pulses.
For different signs of $D_x$ and $D_y$, the ASK is lowered in energy and is the more stable configuration.\\
\section{Conclusions}\label{SecConclusion}
We have introduced a setup that allows for the controlled creation, deletion and propulsion of skyrmions and antiskyrmions in the same sample. 
Both magnetic species can be stabilized in a system with a sufficiently  anisotropic DMI. The creation occurs by a rotation of the magnetization at a part of the edge where the sense of rotation defines the created species. 
By alternating the edge rotation and additionally applying current pulses, an array of skyrmions and antiskyrmions can be created and moved along the sample. A deletion process can be realized by a rotation of opposite sense compared to the creation process as long as the according particle is close to the rotating edge and the racetrack width $N_y$ is not too large.
We could show that the concept is robust against deviations from a strictly one-dimensional DMI allowing for an implementation in existing materials. The controlled creation, deletion and propulsion of skyrmions and antiskyrmions yields the possibility of a racetrack device relying on two different species of quasiparticles. This is in contrast to previously proposed hole-particle designs \cite{fert2013skyrmions} where information is encoded only in the presence and absence of one data carrier. Thus, the here presented racetrack design is more robust against positional fluctuations of the data carriers and could be more densely packed.\\
\section{Acknowledgements}
PS, MT and RW acknowledge funding by the Cluster of Excellence “CUI: Advanced Imaging of Matter” of the Deutsche Forschungsgemeinschaft (DFG)—EXC 2056— project ID 390715994.
MS acknowledges funding by the DFG (project no. 403505707). AS acknowledged financial support from the German research foundation (DFG) through the collaborative research center CRC/TRR 227. TP acknowledges funding by the DFG (project no. 420120155).

\begin{thebibliography}{51}%
\makeatletter
\providecommand \@ifxundefined [1]{%
 \@ifx{#1\undefined}
}%
\providecommand \@ifnum [1]{%
 \ifnum #1\expandafter \@firstoftwo
 \else \expandafter \@secondoftwo
 \fi
}%
\providecommand \@ifx [1]{%
 \ifx #1\expandafter \@firstoftwo
 \else \expandafter \@secondoftwo
 \fi
}%
\providecommand \natexlab [1]{#1}%
\providecommand \enquote  [1]{``#1''}%
\providecommand \bibnamefont  [1]{#1}%
\providecommand \bibfnamefont [1]{#1}%
\providecommand \citenamefont [1]{#1}%
\providecommand \href@noop [0]{\@secondoftwo}%
\providecommand \href [0]{\begingroup \@sanitize@url \@href}%
\providecommand \@href[1]{\@@startlink{#1}\@@href}%
\providecommand \@@href[1]{\endgroup#1\@@endlink}%
\providecommand \@sanitize@url [0]{\catcode `\\12\catcode `\$12\catcode
  `\&12\catcode `\#12\catcode `\^12\catcode `\_12\catcode `\%12\relax}%
\providecommand \@@startlink[1]{}%
\providecommand \@@endlink[0]{}%
\providecommand \url  [0]{\begingroup\@sanitize@url \@url }%
\providecommand \@url [1]{\endgroup\@href {#1}{\urlprefix }}%
\providecommand \urlprefix  [0]{URL }%
\providecommand \Eprint [0]{\href }%
\providecommand \doibase [0]{https://doi.org/}%
\providecommand \selectlanguage [0]{\@gobble}%
\providecommand \bibinfo  [0]{\@secondoftwo}%
\providecommand \bibfield  [0]{\@secondoftwo}%
\providecommand \translation [1]{[#1]}%
\providecommand \BibitemOpen [0]{}%
\providecommand \bibitemStop [0]{}%
\providecommand \bibitemNoStop [0]{.\EOS\space}%
\providecommand \EOS [0]{\spacefactor3000\relax}%
\providecommand \BibitemShut  [1]{\csname bibitem#1\endcsname}%
\let\auto@bib@innerbib\@empty
\bibitem [{\citenamefont {Back}\ \emph {et~al.}(2020)\citenamefont {Back},
  \citenamefont {Cros}, \citenamefont {Ebert}, \citenamefont {Everschor-Sitte},
  \citenamefont {Fert}, \citenamefont {Garst}, \citenamefont {Ma},
  \citenamefont {Mankovsky}, \citenamefont {Monchesky}, \citenamefont
  {Mostovoy}, \citenamefont {Nagaosa}, \citenamefont {Parkin}, \citenamefont
  {Pfleiderer}, \citenamefont {Reyren}, \citenamefont {Rosch}, \citenamefont
  {Taguchi}, \citenamefont {Tokura}, \citenamefont {von Bergmann},\ and\
  \citenamefont {Zang}}]{roadmap}%
  \BibitemOpen
  \bibfield  {author} {\bibinfo {author} {\bibfnamefont {C.}~\bibnamefont
  {Back}}, \bibinfo {author} {\bibfnamefont {V.}~\bibnamefont {Cros}}, \bibinfo
  {author} {\bibfnamefont {H.}~\bibnamefont {Ebert}}, \bibinfo {author}
  {\bibfnamefont {K.}~\bibnamefont {Everschor-Sitte}}, \bibinfo {author}
  {\bibfnamefont {A.}~\bibnamefont {Fert}}, \bibinfo {author} {\bibfnamefont
  {M.}~\bibnamefont {Garst}}, \bibinfo {author} {\bibfnamefont
  {T.}~\bibnamefont {Ma}}, \bibinfo {author} {\bibfnamefont {S.}~\bibnamefont
  {Mankovsky}}, \bibinfo {author} {\bibfnamefont {T.~L.}\ \bibnamefont
  {Monchesky}}, \bibinfo {author} {\bibfnamefont {M.}~\bibnamefont {Mostovoy}},
  \bibinfo {author} {\bibfnamefont {N.}~\bibnamefont {Nagaosa}}, \bibinfo
  {author} {\bibfnamefont {S.~S.~P.}\ \bibnamefont {Parkin}}, \bibinfo {author}
  {\bibfnamefont {C.}~\bibnamefont {Pfleiderer}}, \bibinfo {author}
  {\bibfnamefont {N.}~\bibnamefont {Reyren}}, \bibinfo {author} {\bibfnamefont
  {A.}~\bibnamefont {Rosch}}, \bibinfo {author} {\bibfnamefont
  {Y.}~\bibnamefont {Taguchi}}, \bibinfo {author} {\bibfnamefont
  {Y.}~\bibnamefont {Tokura}}, \bibinfo {author} {\bibfnamefont
  {K.}~\bibnamefont {von Bergmann}},\ and\ \bibinfo {author} {\bibfnamefont
  {J.}~\bibnamefont {Zang}},\ }\bibfield  {title} {\bibinfo {title} {Skyrmions
  and antiskyrmions in quasi-two-dimensional magnets},\ }\href@noop {}
  {\bibfield  {journal} {\bibinfo  {journal} {J. Phys. D: Appl. Phys.}\
  }\textbf {\bibinfo {volume} {53}},\ \bibinfo {pages} {363001} (\bibinfo
  {year} {2020})}\BibitemShut {NoStop}%
\bibitem [{\citenamefont {Fert}\ \emph
  {et~al.}(2013{\natexlab{a}})\citenamefont {Fert}, \citenamefont {Cros},\ and\
  \citenamefont {Sampaio}}]{fert2013skyrmions}%
  \BibitemOpen
  \bibfield  {author} {\bibinfo {author} {\bibfnamefont {A.}~\bibnamefont
  {Fert}}, \bibinfo {author} {\bibfnamefont {V.}~\bibnamefont {Cros}},\ and\
  \bibinfo {author} {\bibfnamefont {J.}~\bibnamefont {Sampaio}},\ }\bibfield
  {title} {\bibinfo {title} {Skyrmions on the track},\ }\href
  {https://www.nature.com/articles/nnano.2013.29} {\bibfield  {journal}
  {\bibinfo  {journal} {Nat. Nanotechnol.}\ }\textbf {\bibinfo {volume} {8}},\
  \bibinfo {pages} {152} (\bibinfo {year} {2013}{\natexlab{a}})}\BibitemShut
  {NoStop}%
\bibitem [{\citenamefont {M{\"u}ller}(2017)}]{muller2017magnetic}%
  \BibitemOpen
  \bibfield  {author} {\bibinfo {author} {\bibfnamefont {J.}~\bibnamefont
  {M{\"u}ller}},\ }\bibfield  {title} {\bibinfo {title} {Magnetic skyrmions on
  a two-lane racetrack},\ }\href
  {https://iopscience.iop.org/article/10.1088/1367-2630/aa5b55} {\bibfield
  {journal} {\bibinfo  {journal} {New J. Phys.}\ }\textbf {\bibinfo {volume}
  {19}},\ \bibinfo {pages} {025002} (\bibinfo {year} {2017})}\BibitemShut
  {NoStop}%
\bibitem [{\citenamefont {Kovalev}\ and\ \citenamefont
  {Sandhoefner}(2018)}]{kovalev2018skyrmions}%
  \BibitemOpen
  \bibfield  {author} {\bibinfo {author} {\bibfnamefont {A.~A.}\ \bibnamefont
  {Kovalev}}\ and\ \bibinfo {author} {\bibfnamefont {S.}~\bibnamefont
  {Sandhoefner}},\ }\bibfield  {title} {\bibinfo {title} {Skyrmions and
  antiskyrmions in quasi-two-dimensional magnets},\ }\href
  {https://www.frontiersin.org/articles/10.3389/fphy.2018.00098/full}
  {\bibfield  {journal} {\bibinfo  {journal} {Front. Phys.}\ }\textbf {\bibinfo
  {volume} {6}},\ \bibinfo {pages} {98} (\bibinfo {year} {2018})}\BibitemShut
  {NoStop}%
\bibitem [{\citenamefont {Nayak}\ \emph {et~al.}(2017)\citenamefont {Nayak},
  \citenamefont {Kumar}, \citenamefont {Ma}, \citenamefont {Werner},
  \citenamefont {Pippel}, \citenamefont {Sahoo}, \citenamefont {Damay},
  \citenamefont {R{\"o}{\ss}ler}, \citenamefont {Felser},\ and\ \citenamefont
  {Parkin}}]{nayak2017magnetic}%
  \BibitemOpen
  \bibfield  {author} {\bibinfo {author} {\bibfnamefont {A.~K.}\ \bibnamefont
  {Nayak}}, \bibinfo {author} {\bibfnamefont {V.}~\bibnamefont {Kumar}},
  \bibinfo {author} {\bibfnamefont {T.}~\bibnamefont {Ma}}, \bibinfo {author}
  {\bibfnamefont {P.}~\bibnamefont {Werner}}, \bibinfo {author} {\bibfnamefont
  {E.}~\bibnamefont {Pippel}}, \bibinfo {author} {\bibfnamefont
  {R.}~\bibnamefont {Sahoo}}, \bibinfo {author} {\bibfnamefont
  {F.}~\bibnamefont {Damay}}, \bibinfo {author} {\bibfnamefont {U.~K.}\
  \bibnamefont {R{\"o}{\ss}ler}}, \bibinfo {author} {\bibfnamefont
  {C.}~\bibnamefont {Felser}},\ and\ \bibinfo {author} {\bibfnamefont {S.~S.}\
  \bibnamefont {Parkin}},\ }\bibfield  {title} {\bibinfo {title} {Magnetic
  antiskyrmions above room temperature in tetragonal heusler materials},\
  }\href {https://www.nature.com/articles/nature23466/} {\bibfield  {journal}
  {\bibinfo  {journal} {Nature}\ }\textbf {\bibinfo {volume} {548}},\ \bibinfo
  {pages} {561} (\bibinfo {year} {2017})}\BibitemShut {NoStop}%
\bibitem [{\citenamefont {Jena}\ \emph {et~al.}(2019)\citenamefont {Jena},
  \citenamefont {Stinshoff}, \citenamefont {Saha}, \citenamefont {Srivastava},
  \citenamefont {Ma}, \citenamefont {Deniz}, \citenamefont {Werner},
  \citenamefont {Felser},\ and\ \citenamefont {Parkin}}]{jena2019observation}%
  \BibitemOpen
  \bibfield  {author} {\bibinfo {author} {\bibfnamefont {J.}~\bibnamefont
  {Jena}}, \bibinfo {author} {\bibfnamefont {R.}~\bibnamefont {Stinshoff}},
  \bibinfo {author} {\bibfnamefont {R.}~\bibnamefont {Saha}}, \bibinfo {author}
  {\bibfnamefont {A.~K.}\ \bibnamefont {Srivastava}}, \bibinfo {author}
  {\bibfnamefont {T.}~\bibnamefont {Ma}}, \bibinfo {author} {\bibfnamefont
  {H.}~\bibnamefont {Deniz}}, \bibinfo {author} {\bibfnamefont
  {P.}~\bibnamefont {Werner}}, \bibinfo {author} {\bibfnamefont
  {C.}~\bibnamefont {Felser}},\ and\ \bibinfo {author} {\bibfnamefont {S.~S.}\
  \bibnamefont {Parkin}},\ }\bibfield  {title} {\bibinfo {title} {Observation
  of magnetic antiskyrmions in the low magnetization ferrimagnet
  $\text{Mn}_2\text{Rh}_{0. 95}\text{Ir}_{0. 05}\text{Sn}$},\ }\href
  {https://pubs.acs.org/doi/10.1021/acs.nanolett.9b02973} {\bibfield  {journal}
  {\bibinfo  {journal} {Nano Lett.}\ }\textbf {\bibinfo {volume} {20}},\
  \bibinfo {pages} {59} (\bibinfo {year} {2019})}\BibitemShut {NoStop}%
\bibitem [{\citenamefont {Everschor-Sitte}\ \emph {et~al.}(2018)\citenamefont
  {Everschor-Sitte}, \citenamefont {Masell}, \citenamefont {Reeve},\ and\
  \citenamefont {Kl{\"a}ui}}]{everschor2018perspective}%
  \BibitemOpen
  \bibfield  {author} {\bibinfo {author} {\bibfnamefont {K.}~\bibnamefont
  {Everschor-Sitte}}, \bibinfo {author} {\bibfnamefont {J.}~\bibnamefont
  {Masell}}, \bibinfo {author} {\bibfnamefont {R.~M.}\ \bibnamefont {Reeve}},\
  and\ \bibinfo {author} {\bibfnamefont {M.}~\bibnamefont {Kl{\"a}ui}},\
  }\bibfield  {title} {\bibinfo {title} {Perspective: Magnetic
  skyrmions—overview of recent progress in an active research field},\ }\href
  {https://aip.scitation.org/doi/10.1063/1.5048972} {\bibfield  {journal}
  {\bibinfo  {journal} {J. Appl. Phys.}\ }\textbf {\bibinfo {volume} {124}},\
  \bibinfo {pages} {240901} (\bibinfo {year} {2018})}\BibitemShut {NoStop}%
\bibitem [{\citenamefont {Nagaosa}\ and\ \citenamefont
  {Tokura}(2013)}]{Nagaosa2013}%
  \BibitemOpen
  \bibfield  {author} {\bibinfo {author} {\bibfnamefont {N.}~\bibnamefont
  {Nagaosa}}\ and\ \bibinfo {author} {\bibfnamefont {Y.}~\bibnamefont
  {Tokura}},\ }\bibfield  {title} {\bibinfo {title} {Topological properties and
  dynamics of magnetic skyrmions},\ }\href
  {https://doi.org/10.1038/nnano.2013.243} {\bibfield  {journal} {\bibinfo
  {journal} {Nat. Nanotechnol.}\ }\textbf {\bibinfo {volume} {8}},\ \bibinfo
  {pages} {899} (\bibinfo {year} {2013})}\BibitemShut {NoStop}%
\bibitem [{\citenamefont {Vedmedenko}\ and\ \citenamefont
  {Wiesendanger}(2019)}]{Vedmedenko2019}%
  \BibitemOpen
  \bibfield  {author} {\bibinfo {author} {\bibfnamefont {E.}~\bibnamefont
  {Vedmedenko}}\ and\ \bibinfo {author} {\bibfnamefont {R.}~\bibnamefont
  {Wiesendanger}},\ }\bibinfo {title} {Magnetic skyrmions on discrete
  lattices},\ in\ \href@noop {} {\emph {\bibinfo {booktitle} {Spintronic
  Handbook: Spin Transport and Magnetism}}},\ \bibinfo {editor} {edited by\
  \bibinfo {editor} {\bibfnamefont {I.~Z.}\ \bibnamefont {Evgeny Y.~Tsymbal}}}\
  (\bibinfo  {publisher} {CRC Press, Taylor \& Francis},\ \bibinfo {address}
  {Boca Raton},\ \bibinfo {year} {2019})\ Chap.~\bibinfo {chapter} {10}, pp.\
  \bibinfo {pages} {323--357},\ \bibinfo {edition} {2nd}\ ed.\BibitemShut
  {Stop}%
\bibitem [{\citenamefont {Lobanov}\ \emph {et~al.}(2016)\citenamefont
  {Lobanov}, \citenamefont {J\'onsson},\ and\ \citenamefont
  {Uzdin}}]{PhysRevB.94.174418}%
  \BibitemOpen
  \bibfield  {author} {\bibinfo {author} {\bibfnamefont {I.~S.}\ \bibnamefont
  {Lobanov}}, \bibinfo {author} {\bibfnamefont {H.}~\bibnamefont {J\'onsson}},\
  and\ \bibinfo {author} {\bibfnamefont {V.~M.}\ \bibnamefont {Uzdin}},\
  }\bibfield  {title} {\bibinfo {title} {Mechanism and activation energy of
  magnetic skyrmion annihilation obtained from minimum energy path
  calculations},\ }\href {https://doi.org/10.1103/PhysRevB.94.174418}
  {\bibfield  {journal} {\bibinfo  {journal} {Phys. Rev. B}\ }\textbf {\bibinfo
  {volume} {94}},\ \bibinfo {pages} {174418} (\bibinfo {year}
  {2016})}\BibitemShut {NoStop}%
\bibitem [{\citenamefont {Lin}\ \emph {et~al.}(2013)\citenamefont {Lin},
  \citenamefont {Reichhardt},\ and\ \citenamefont
  {Saxena}}]{lin2013manipulation}%
  \BibitemOpen
  \bibfield  {author} {\bibinfo {author} {\bibfnamefont {S.-Z.}\ \bibnamefont
  {Lin}}, \bibinfo {author} {\bibfnamefont {C.}~\bibnamefont {Reichhardt}},\
  and\ \bibinfo {author} {\bibfnamefont {A.}~\bibnamefont {Saxena}},\
  }\bibfield  {title} {\bibinfo {title} {Manipulation of skyrmions in nanodisks
  with a current pulse and skyrmion rectifier},\ }\href
  {https://aip.scitation.org/doi/10.1063/1.4809751} {\bibfield  {journal}
  {\bibinfo  {journal} {Appl. Phys. Lett.}\ }\textbf {\bibinfo {volume}
  {102}},\ \bibinfo {pages} {222405} (\bibinfo {year} {2013})}\BibitemShut
  {NoStop}%
\bibitem [{\citenamefont {Romming}\ \emph {et~al.}(2013)\citenamefont
  {Romming}, \citenamefont {Hanneken}, \citenamefont {Menzel}, \citenamefont
  {Bickel}, \citenamefont {Wolter}, \citenamefont {von Bergmann}, \citenamefont
  {Kubetzka},\ and\ \citenamefont {Wiesendanger}}]{romming2013writing}%
  \BibitemOpen
  \bibfield  {author} {\bibinfo {author} {\bibfnamefont {N.}~\bibnamefont
  {Romming}}, \bibinfo {author} {\bibfnamefont {C.}~\bibnamefont {Hanneken}},
  \bibinfo {author} {\bibfnamefont {M.}~\bibnamefont {Menzel}}, \bibinfo
  {author} {\bibfnamefont {J.~E.}\ \bibnamefont {Bickel}}, \bibinfo {author}
  {\bibfnamefont {B.}~\bibnamefont {Wolter}}, \bibinfo {author} {\bibfnamefont
  {K.}~\bibnamefont {von Bergmann}}, \bibinfo {author} {\bibfnamefont
  {A.}~\bibnamefont {Kubetzka}},\ and\ \bibinfo {author} {\bibfnamefont
  {R.}~\bibnamefont {Wiesendanger}},\ }\bibfield  {title} {\bibinfo {title}
  {Writing and deleting single magnetic skyrmions},\ }\href
  {https://www.science.org/doi/10.1126/science.1240573} {\bibfield  {journal}
  {\bibinfo  {journal} {Science}\ }\textbf {\bibinfo {volume} {341}},\ \bibinfo
  {pages} {636} (\bibinfo {year} {2013})}\BibitemShut {NoStop}%
\bibitem [{\citenamefont {Hsu}\ \emph {et~al.}(2017)\citenamefont {Hsu},
  \citenamefont {Kubetzka}, \citenamefont {Finco}, \citenamefont {Romming},
  \citenamefont {von Bergmann},\ and\ \citenamefont {Wiesendanger}}]{Hsu2017}%
  \BibitemOpen
  \bibfield  {author} {\bibinfo {author} {\bibfnamefont {P.-J.}\ \bibnamefont
  {Hsu}}, \bibinfo {author} {\bibfnamefont {A.}~\bibnamefont {Kubetzka}},
  \bibinfo {author} {\bibfnamefont {A.}~\bibnamefont {Finco}}, \bibinfo
  {author} {\bibfnamefont {N.}~\bibnamefont {Romming}}, \bibinfo {author}
  {\bibfnamefont {K.}~\bibnamefont {von Bergmann}},\ and\ \bibinfo {author}
  {\bibfnamefont {R.}~\bibnamefont {Wiesendanger}},\ }\bibfield  {title}
  {\bibinfo {title} {Electric-field-driven switching of individual magnetic
  skyrmions},\ }\href {https://doi.org/10.1038/nnano.2016.234} {\bibfield
  {journal} {\bibinfo  {journal} {Nat. Nanotechnol.}\ }\textbf {\bibinfo
  {volume} {12}},\ \bibinfo {pages} {123} (\bibinfo {year} {2017})}\BibitemShut
  {NoStop}%
\bibitem [{\citenamefont {Huang}\ \emph {et~al.}(2018)\citenamefont {Huang},
  \citenamefont {Cantoni}, \citenamefont {Kruchkov}, \citenamefont {Rajeswari},
  \citenamefont {Magrez}, \citenamefont {Carbone},\ and\ \citenamefont
  {Rønnow}}]{electricfieldinduced}%
  \BibitemOpen
  \bibfield  {author} {\bibinfo {author} {\bibfnamefont {P.}~\bibnamefont
  {Huang}}, \bibinfo {author} {\bibfnamefont {M.}~\bibnamefont {Cantoni}},
  \bibinfo {author} {\bibfnamefont {A.}~\bibnamefont {Kruchkov}}, \bibinfo
  {author} {\bibfnamefont {J.}~\bibnamefont {Rajeswari}}, \bibinfo {author}
  {\bibfnamefont {A.}~\bibnamefont {Magrez}}, \bibinfo {author} {\bibfnamefont
  {F.}~\bibnamefont {Carbone}},\ and\ \bibinfo {author} {\bibfnamefont {H.~M.}\
  \bibnamefont {Rønnow}},\ }\bibfield  {title} {\bibinfo {title} {In situ
  electric field skyrmion creation in magnetoelectric $cu_2oseo_3$},\ }\href
  {https://doi.org/10.1021/acs.nanolett.8b02097} {\bibfield  {journal}
  {\bibinfo  {journal} {Nano Lett.}\ }\textbf {\bibinfo {volume} {18}},\
  \bibinfo {pages} {5167} (\bibinfo {year} {2018})}\BibitemShut {NoStop}%
\bibitem [{\citenamefont {Sch{\"{a}}ffer}\ \emph {et~al.}(2020)\citenamefont
  {Sch{\"{a}}ffer}, \citenamefont {Siegl}, \citenamefont {Stier}, \citenamefont
  {Posske}, \citenamefont {Berakdar}, \citenamefont {Thorwart}, \citenamefont
  {Wiesendanger},\ and\ \citenamefont {Vedmedenko}}]{Schaffer2020}%
  \BibitemOpen
  \bibfield  {author} {\bibinfo {author} {\bibfnamefont {A.~F.}\ \bibnamefont
  {Sch{\"{a}}ffer}}, \bibinfo {author} {\bibfnamefont {P.}~\bibnamefont
  {Siegl}}, \bibinfo {author} {\bibfnamefont {M.}~\bibnamefont {Stier}},
  \bibinfo {author} {\bibfnamefont {T.}~\bibnamefont {Posske}}, \bibinfo
  {author} {\bibfnamefont {J.}~\bibnamefont {Berakdar}}, \bibinfo {author}
  {\bibfnamefont {M.}~\bibnamefont {Thorwart}}, \bibinfo {author}
  {\bibfnamefont {R.}~\bibnamefont {Wiesendanger}},\ and\ \bibinfo {author}
  {\bibfnamefont {E.~Y.}\ \bibnamefont {Vedmedenko}},\ }\bibfield  {title}
  {\bibinfo {title} {{Rotating edge-field driven processing of chiral spin
  textures in racetrack devices}},\ }\href
  {https://doi.org/10.1038/s41598-020-77337-y} {\bibfield  {journal} {\bibinfo
  {journal} {Sci. Rep.}\ }\textbf {\bibinfo {volume} {10}},\ \bibinfo {pages}
  {1} (\bibinfo {year} {2020})}\BibitemShut {NoStop}%
\bibitem [{\citenamefont {Stier}\ \emph {et~al.}(2017)\citenamefont {Stier},
  \citenamefont {H{\"a}usler}, \citenamefont {Posske}, \citenamefont {Gurski},\
  and\ \citenamefont {Thorwart}}]{stier2017skyrmion}%
  \BibitemOpen
  \bibfield  {author} {\bibinfo {author} {\bibfnamefont {M.}~\bibnamefont
  {Stier}}, \bibinfo {author} {\bibfnamefont {W.}~\bibnamefont {H{\"a}usler}},
  \bibinfo {author} {\bibfnamefont {T.}~\bibnamefont {Posske}}, \bibinfo
  {author} {\bibfnamefont {G.}~\bibnamefont {Gurski}},\ and\ \bibinfo {author}
  {\bibfnamefont {M.}~\bibnamefont {Thorwart}},\ }\bibfield  {title} {\bibinfo
  {title} {Skyrmion--anti-skyrmion pair creation by in-plane currents},\ }\href
  {https://journals.aps.org/prl/abstract/10.1103/PhysRevLett.118.267203}
  {\bibfield  {journal} {\bibinfo  {journal} {Phys. Rev. Lett.}\ }\textbf
  {\bibinfo {volume} {118}},\ \bibinfo {pages} {267203} (\bibinfo {year}
  {2017})}\BibitemShut {NoStop}%
\bibitem [{\citenamefont {Everschor-Sitte}\ \emph {et~al.}(2017)\citenamefont
  {Everschor-Sitte}, \citenamefont {Sitte}, \citenamefont {Valet},
  \citenamefont {Abanov},\ and\ \citenamefont
  {Sinova}}]{everschor2017skyrmion}%
  \BibitemOpen
  \bibfield  {author} {\bibinfo {author} {\bibfnamefont {K.}~\bibnamefont
  {Everschor-Sitte}}, \bibinfo {author} {\bibfnamefont {M.}~\bibnamefont
  {Sitte}}, \bibinfo {author} {\bibfnamefont {T.}~\bibnamefont {Valet}},
  \bibinfo {author} {\bibfnamefont {A.}~\bibnamefont {Abanov}},\ and\ \bibinfo
  {author} {\bibfnamefont {J.}~\bibnamefont {Sinova}},\ }\bibfield  {title}
  {\bibinfo {title} {Skyrmion production on demand by homogeneous dc
  currents},\ }\href
  {https://iopscience.iop.org/article/10.1088/1367-2630/aa8569} {\bibfield
  {journal} {\bibinfo  {journal} {New J. Phys.}\ }\textbf {\bibinfo {volume}
  {19}},\ \bibinfo {pages} {092001} (\bibinfo {year} {2017})}\BibitemShut
  {NoStop}%
\bibitem [{\citenamefont {Yuan}\ and\ \citenamefont {Wang}(2016)}]{Yuan2016}%
  \BibitemOpen
  \bibfield  {author} {\bibinfo {author} {\bibfnamefont {H.~Y.}\ \bibnamefont
  {Yuan}}\ and\ \bibinfo {author} {\bibfnamefont {X.~R.}\ \bibnamefont
  {Wang}},\ }\bibfield  {title} {\bibinfo {title} {Skyrmion creation and
  manipulation by nano-second current pulses},\ }\href
  {https://doi.org/10.1038/srep22638} {\bibfield  {journal} {\bibinfo
  {journal} {Sci. Rep.}\ }\textbf {\bibinfo {volume} {6}},\ \bibinfo {pages}
  {22638} (\bibinfo {year} {2016})}\BibitemShut {NoStop}%
\bibitem [{\citenamefont {Liu}\ \emph {et~al.}(2015)\citenamefont {Liu},
  \citenamefont {Yin}, \citenamefont {Zang}, \citenamefont {Shi},\ and\
  \citenamefont {Lake}}]{spinwavecreation2015}%
  \BibitemOpen
  \bibfield  {author} {\bibinfo {author} {\bibfnamefont {Y.}~\bibnamefont
  {Liu}}, \bibinfo {author} {\bibfnamefont {G.}~\bibnamefont {Yin}}, \bibinfo
  {author} {\bibfnamefont {J.}~\bibnamefont {Zang}}, \bibinfo {author}
  {\bibfnamefont {J.}~\bibnamefont {Shi}},\ and\ \bibinfo {author}
  {\bibfnamefont {R.~K.}\ \bibnamefont {Lake}},\ }\bibfield  {title} {\bibinfo
  {title} {Skyrmion creation and annihilation by spin waves},\ }\href
  {https://doi.org/10.1063/1.4933407} {\bibfield  {journal} {\bibinfo
  {journal} {Appl. Phys. Lett.}\ }\textbf {\bibinfo {volume} {107}},\ \bibinfo
  {pages} {152411} (\bibinfo {year} {2015})}\BibitemShut {NoStop}%
\bibitem [{\citenamefont {Je}\ \emph {et~al.}(2018)\citenamefont {Je},
  \citenamefont {Vallobra}, \citenamefont {Srivastava}, \citenamefont
  {Rojas-Sánchez}, \citenamefont {Pham}, \citenamefont {Hehn}, \citenamefont
  {Malinowski}, \citenamefont {Baraduc}, \citenamefont {Auffret}, \citenamefont
  {Gaudin}, \citenamefont {Mangin}, \citenamefont {Béa},\ and\ \citenamefont
  {Boulle}}]{laserinducedcreation}%
  \BibitemOpen
  \bibfield  {author} {\bibinfo {author} {\bibfnamefont {S.-G.}\ \bibnamefont
  {Je}}, \bibinfo {author} {\bibfnamefont {P.}~\bibnamefont {Vallobra}},
  \bibinfo {author} {\bibfnamefont {T.}~\bibnamefont {Srivastava}}, \bibinfo
  {author} {\bibfnamefont {J.-C.}\ \bibnamefont {Rojas-Sánchez}}, \bibinfo
  {author} {\bibfnamefont {T.~H.}\ \bibnamefont {Pham}}, \bibinfo {author}
  {\bibfnamefont {M.}~\bibnamefont {Hehn}}, \bibinfo {author} {\bibfnamefont
  {G.}~\bibnamefont {Malinowski}}, \bibinfo {author} {\bibfnamefont
  {C.}~\bibnamefont {Baraduc}}, \bibinfo {author} {\bibfnamefont
  {S.}~\bibnamefont {Auffret}}, \bibinfo {author} {\bibfnamefont
  {G.}~\bibnamefont {Gaudin}}, \bibinfo {author} {\bibfnamefont
  {S.}~\bibnamefont {Mangin}}, \bibinfo {author} {\bibfnamefont
  {H.}~\bibnamefont {Béa}},\ and\ \bibinfo {author} {\bibfnamefont
  {O.}~\bibnamefont {Boulle}},\ }\bibfield  {title} {\bibinfo {title} {Creation
  of magnetic skyrmion bubble lattices by ultrafast laser in ultrathin films},\
  }\href {https://doi.org/10.1021/acs.nanolett.8b03653} {\bibfield  {journal}
  {\bibinfo  {journal} {Nano Lett.}\ }\textbf {\bibinfo {volume} {18}},\
  \bibinfo {pages} {7362} (\bibinfo {year} {2018})}\BibitemShut {NoStop}%
\bibitem [{\citenamefont {Berruto}\ \emph {et~al.}(2018)\citenamefont
  {Berruto}, \citenamefont {Madan}, \citenamefont {Murooka}, \citenamefont
  {Vanacore}, \citenamefont {Pomarico}, \citenamefont {Rajeswari},
  \citenamefont {Lamb}, \citenamefont {Huang}, \citenamefont {Kruchkov},
  \citenamefont {Togawa}, \citenamefont {LaGrange}, \citenamefont {McGrouther},
  \citenamefont {R\o{}nnow},\ and\ \citenamefont
  {Carbone}}]{laserinducedcreation2}%
  \BibitemOpen
  \bibfield  {author} {\bibinfo {author} {\bibfnamefont {G.}~\bibnamefont
  {Berruto}}, \bibinfo {author} {\bibfnamefont {I.}~\bibnamefont {Madan}},
  \bibinfo {author} {\bibfnamefont {Y.}~\bibnamefont {Murooka}}, \bibinfo
  {author} {\bibfnamefont {G.~M.}\ \bibnamefont {Vanacore}}, \bibinfo {author}
  {\bibfnamefont {E.}~\bibnamefont {Pomarico}}, \bibinfo {author}
  {\bibfnamefont {J.}~\bibnamefont {Rajeswari}}, \bibinfo {author}
  {\bibfnamefont {R.}~\bibnamefont {Lamb}}, \bibinfo {author} {\bibfnamefont
  {P.}~\bibnamefont {Huang}}, \bibinfo {author} {\bibfnamefont {A.~J.}\
  \bibnamefont {Kruchkov}}, \bibinfo {author} {\bibfnamefont {Y.}~\bibnamefont
  {Togawa}}, \bibinfo {author} {\bibfnamefont {T.}~\bibnamefont {LaGrange}},
  \bibinfo {author} {\bibfnamefont {D.}~\bibnamefont {McGrouther}}, \bibinfo
  {author} {\bibfnamefont {H.~M.}\ \bibnamefont {R\o{}nnow}},\ and\ \bibinfo
  {author} {\bibfnamefont {F.}~\bibnamefont {Carbone}},\ }\bibfield  {title}
  {\bibinfo {title} {Laser-induced skyrmion writing and erasing in an ultrafast
  cryo-lorentz transmission electron microscope},\ }\href
  {https://doi.org/10.1103/PhysRevLett.120.117201} {\bibfield  {journal}
  {\bibinfo  {journal} {Phys. Rev. Lett.}\ }\textbf {\bibinfo {volume} {120}},\
  \bibinfo {pages} {117201} (\bibinfo {year} {2018})}\BibitemShut {NoStop}%
\bibitem [{\citenamefont {Koshibae}\ and\ \citenamefont
  {Nagaosa}(2014)}]{koshibae2014creation}%
  \BibitemOpen
  \bibfield  {author} {\bibinfo {author} {\bibfnamefont {W.}~\bibnamefont
  {Koshibae}}\ and\ \bibinfo {author} {\bibfnamefont {N.}~\bibnamefont
  {Nagaosa}},\ }\bibfield  {title} {\bibinfo {title} {Creation of skyrmions and
  antiskyrmions by local heating},\ }\href
  {https://www.nature.com/articles/ncomms6148} {\bibfield  {journal} {\bibinfo
  {journal} {Nat. Commun.}\ }\textbf {\bibinfo {volume} {5}},\ \bibinfo {pages}
  {1} (\bibinfo {year} {2014})}\BibitemShut {NoStop}%
\bibitem [{\citenamefont {Jiang}\ \emph {et~al.}(2015)\citenamefont {Jiang},
  \citenamefont {Upadhyaya}, \citenamefont {Zhang}, \citenamefont {Yu},
  \citenamefont {Jungfleisch}, \citenamefont {Fradin}, \citenamefont {Pearson},
  \citenamefont {Tserkovnyak}, \citenamefont {Wang}, \citenamefont {Heinonen}
  \emph {et~al.}}]{jiang2015blowing}%
  \BibitemOpen
  \bibfield  {author} {\bibinfo {author} {\bibfnamefont {W.}~\bibnamefont
  {Jiang}}, \bibinfo {author} {\bibfnamefont {P.}~\bibnamefont {Upadhyaya}},
  \bibinfo {author} {\bibfnamefont {W.}~\bibnamefont {Zhang}}, \bibinfo
  {author} {\bibfnamefont {G.}~\bibnamefont {Yu}}, \bibinfo {author}
  {\bibfnamefont {M.~B.}\ \bibnamefont {Jungfleisch}}, \bibinfo {author}
  {\bibfnamefont {F.~Y.}\ \bibnamefont {Fradin}}, \bibinfo {author}
  {\bibfnamefont {J.~E.}\ \bibnamefont {Pearson}}, \bibinfo {author}
  {\bibfnamefont {Y.}~\bibnamefont {Tserkovnyak}}, \bibinfo {author}
  {\bibfnamefont {K.~L.}\ \bibnamefont {Wang}}, \bibinfo {author}
  {\bibfnamefont {O.}~\bibnamefont {Heinonen}}, \emph {et~al.},\ }\bibfield
  {title} {\bibinfo {title} {Blowing magnetic skyrmion bubbles},\ }\href
  {https://www.science.org/doi/10.1126/science.aaa1442} {\bibfield  {journal}
  {\bibinfo  {journal} {Science}\ }\textbf {\bibinfo {volume} {349}},\ \bibinfo
  {pages} {283} (\bibinfo {year} {2015})}\BibitemShut {NoStop}%
\bibitem [{\citenamefont {Iwasaki}\ \emph {et~al.}(2013)\citenamefont
  {Iwasaki}, \citenamefont {Mochizuki},\ and\ \citenamefont
  {Nagaosa}}]{iwasaki2013current}%
  \BibitemOpen
  \bibfield  {author} {\bibinfo {author} {\bibfnamefont {J.}~\bibnamefont
  {Iwasaki}}, \bibinfo {author} {\bibfnamefont {M.}~\bibnamefont {Mochizuki}},\
  and\ \bibinfo {author} {\bibfnamefont {N.}~\bibnamefont {Nagaosa}},\
  }\bibfield  {title} {\bibinfo {title} {Current-induced skyrmion dynamics in
  constricted geometries},\ }\href
  {https://www.nature.com/articles/nnano.2013.176} {\bibfield  {journal}
  {\bibinfo  {journal} {Nat. Nanotechnol.}\ }\textbf {\bibinfo {volume} {8}},\
  \bibinfo {pages} {742} (\bibinfo {year} {2013})}\BibitemShut {NoStop}%
\bibitem [{\citenamefont {Zhang}\ \emph {et~al.}(2018)\citenamefont {Zhang},
  \citenamefont {Wang}, \citenamefont {Burn}, \citenamefont {Peng},
  \citenamefont {Berger}, \citenamefont {Bauer}, \citenamefont {Pfleiderer},
  \citenamefont {van~der Laan},\ and\ \citenamefont {Hesjedal}}]{Zhang2018}%
  \BibitemOpen
  \bibfield  {author} {\bibinfo {author} {\bibfnamefont {S.~L.}\ \bibnamefont
  {Zhang}}, \bibinfo {author} {\bibfnamefont {W.~W.}\ \bibnamefont {Wang}},
  \bibinfo {author} {\bibfnamefont {D.~M.}\ \bibnamefont {Burn}}, \bibinfo
  {author} {\bibfnamefont {H.}~\bibnamefont {Peng}}, \bibinfo {author}
  {\bibfnamefont {H.}~\bibnamefont {Berger}}, \bibinfo {author} {\bibfnamefont
  {A.}~\bibnamefont {Bauer}}, \bibinfo {author} {\bibfnamefont
  {C.}~\bibnamefont {Pfleiderer}}, \bibinfo {author} {\bibfnamefont
  {G.}~\bibnamefont {van~der Laan}},\ and\ \bibinfo {author} {\bibfnamefont
  {T.}~\bibnamefont {Hesjedal}},\ }\bibfield  {title} {\bibinfo {title}
  {Manipulation of skyrmion motion by magnetic field gradients},\ }\href
  {https://doi.org/10.1038/s41467-018-04563-4} {\bibfield  {journal} {\bibinfo
  {journal} {Nat. Commun.}\ }\textbf {\bibinfo {volume} {9}},\ \bibinfo {pages}
  {2115} (\bibinfo {year} {2018})}\BibitemShut {NoStop}%
\bibitem [{\citenamefont {Moon}\ \emph {et~al.}(2016)\citenamefont {Moon},
  \citenamefont {Kim}, \citenamefont {Je}, \citenamefont {Chun}, \citenamefont
  {Kim}, \citenamefont {Qiu}, \citenamefont {Choe},\ and\ \citenamefont
  {Hwang}}]{moon2016skyrmion}%
  \BibitemOpen
  \bibfield  {author} {\bibinfo {author} {\bibfnamefont {K.-W.}\ \bibnamefont
  {Moon}}, \bibinfo {author} {\bibfnamefont {D.-H.}\ \bibnamefont {Kim}},
  \bibinfo {author} {\bibfnamefont {S.-G.}\ \bibnamefont {Je}}, \bibinfo
  {author} {\bibfnamefont {B.~S.}\ \bibnamefont {Chun}}, \bibinfo {author}
  {\bibfnamefont {W.}~\bibnamefont {Kim}}, \bibinfo {author} {\bibfnamefont
  {Z.}~\bibnamefont {Qiu}}, \bibinfo {author} {\bibfnamefont {S.-B.}\
  \bibnamefont {Choe}},\ and\ \bibinfo {author} {\bibfnamefont
  {C.}~\bibnamefont {Hwang}},\ }\bibfield  {title} {\bibinfo {title} {Skyrmion
  motion driven by oscillating magnetic field},\ }\href
  {https://www.nature.com/articles/srep20360} {\bibfield  {journal} {\bibinfo
  {journal} {Sci. Rep.}\ }\textbf {\bibinfo {volume} {6}},\ \bibinfo {pages}
  {1} (\bibinfo {year} {2016})}\BibitemShut {NoStop}%
\bibitem [{\citenamefont {Chen}(2017)}]{chen2017skyrmion}%
  \BibitemOpen
  \bibfield  {author} {\bibinfo {author} {\bibfnamefont {G.}~\bibnamefont
  {Chen}},\ }\bibfield  {title} {\bibinfo {title} {Skyrmion hall effect},\
  }\href {https://www.nature.com/articles/nphys4030} {\bibfield  {journal}
  {\bibinfo  {journal} {Nat. Phys.}\ }\textbf {\bibinfo {volume} {13}},\
  \bibinfo {pages} {112} (\bibinfo {year} {2017})}\BibitemShut {NoStop}%
\bibitem [{\citenamefont {Jiang}\ \emph {et~al.}(2017)\citenamefont {Jiang},
  \citenamefont {Zhang}, \citenamefont {Yu}, \citenamefont {Zhang},
  \citenamefont {Wang}, \citenamefont {Jungfleisch}, \citenamefont {Pearson},
  \citenamefont {Cheng}, \citenamefont {Heinonen}, \citenamefont {Wang} \emph
  {et~al.}}]{jiang2017direct}%
  \BibitemOpen
  \bibfield  {author} {\bibinfo {author} {\bibfnamefont {W.}~\bibnamefont
  {Jiang}}, \bibinfo {author} {\bibfnamefont {X.}~\bibnamefont {Zhang}},
  \bibinfo {author} {\bibfnamefont {G.}~\bibnamefont {Yu}}, \bibinfo {author}
  {\bibfnamefont {W.}~\bibnamefont {Zhang}}, \bibinfo {author} {\bibfnamefont
  {X.}~\bibnamefont {Wang}}, \bibinfo {author} {\bibfnamefont {M.~B.}\
  \bibnamefont {Jungfleisch}}, \bibinfo {author} {\bibfnamefont {J.~E.}\
  \bibnamefont {Pearson}}, \bibinfo {author} {\bibfnamefont {X.}~\bibnamefont
  {Cheng}}, \bibinfo {author} {\bibfnamefont {O.}~\bibnamefont {Heinonen}},
  \bibinfo {author} {\bibfnamefont {K.~L.}\ \bibnamefont {Wang}}, \emph
  {et~al.},\ }\bibfield  {title} {\bibinfo {title} {Direct observation of the
  skyrmion hall effect},\ }\href {https://www.nature.com/articles/nphys3883/}
  {\bibfield  {journal} {\bibinfo  {journal} {Nat. Phys.}\ }\textbf {\bibinfo
  {volume} {13}},\ \bibinfo {pages} {162} (\bibinfo {year} {2017})}\BibitemShut
  {NoStop}%
\bibitem [{\citenamefont {Plettenberg}\ \emph {et~al.}(2020)\citenamefont
  {Plettenberg}, \citenamefont {Stier},\ and\ \citenamefont
  {Thorwart}}]{plettenberg2020steering}%
  \BibitemOpen
  \bibfield  {author} {\bibinfo {author} {\bibfnamefont {J.}~\bibnamefont
  {Plettenberg}}, \bibinfo {author} {\bibfnamefont {M.}~\bibnamefont {Stier}},\
  and\ \bibinfo {author} {\bibfnamefont {M.}~\bibnamefont {Thorwart}},\
  }\bibfield  {title} {\bibinfo {title} {Steering of the skyrmion hall angle by
  gate voltages},\ }\href
  {https://journals.aps.org/prl/abstract/10.1103/PhysRevLett.124.207202}
  {\bibfield  {journal} {\bibinfo  {journal} {Phys. Rev. Lett.}\ }\textbf
  {\bibinfo {volume} {124}},\ \bibinfo {pages} {207202} (\bibinfo {year}
  {2020})}\BibitemShut {NoStop}%
\bibitem [{\citenamefont {Potkina}\ \emph {et~al.}(2020)\citenamefont
  {Potkina}, \citenamefont {Lobanov}, \citenamefont {J{\'o}nsson},\ and\
  \citenamefont {Uzdin}}]{potkina2020skyrmions}%
  \BibitemOpen
  \bibfield  {author} {\bibinfo {author} {\bibfnamefont {M.~N.}\ \bibnamefont
  {Potkina}}, \bibinfo {author} {\bibfnamefont {I.~S.}\ \bibnamefont
  {Lobanov}}, \bibinfo {author} {\bibfnamefont {H.}~\bibnamefont
  {J{\'o}nsson}},\ and\ \bibinfo {author} {\bibfnamefont {V.~M.}\ \bibnamefont
  {Uzdin}},\ }\bibfield  {title} {\bibinfo {title} {Skyrmions in
  antiferromagnets: Thermal stability and the effect of external field and
  impurities},\ }\href {https://aip.scitation.org/doi/10.1063/5.0009559}
  {\bibfield  {journal} {\bibinfo  {journal} {J. Appl. Phys.}\ }\textbf
  {\bibinfo {volume} {127}},\ \bibinfo {pages} {213906} (\bibinfo {year}
  {2020})}\BibitemShut {NoStop}%
\bibitem [{\citenamefont {Stier}\ \emph {et~al.}(2021)\citenamefont {Stier},
  \citenamefont {Strobel}, \citenamefont {Krause}, \citenamefont
  {H{\"a}usler},\ and\ \citenamefont {Thorwart}}]{stier2021role}%
  \BibitemOpen
  \bibfield  {author} {\bibinfo {author} {\bibfnamefont {M.}~\bibnamefont
  {Stier}}, \bibinfo {author} {\bibfnamefont {R.}~\bibnamefont {Strobel}},
  \bibinfo {author} {\bibfnamefont {S.}~\bibnamefont {Krause}}, \bibinfo
  {author} {\bibfnamefont {W.}~\bibnamefont {H{\"a}usler}},\ and\ \bibinfo
  {author} {\bibfnamefont {M.}~\bibnamefont {Thorwart}},\ }\bibfield  {title}
  {\bibinfo {title} {Role of impurity clusters for the current-driven motion of
  magnetic skyrmions},\ }\href
  {https://journals.aps.org/prb/abstract/10.1103/PhysRevB.103.054420}
  {\bibfield  {journal} {\bibinfo  {journal} {Phys. Rev. B}\ }\textbf {\bibinfo
  {volume} {103}},\ \bibinfo {pages} {054420} (\bibinfo {year}
  {2021})}\BibitemShut {NoStop}%
\bibitem [{\citenamefont {Song}\ \emph {et~al.}(2017)\citenamefont {Song},
  \citenamefont {Jin}, \citenamefont {Wang}, \citenamefont {Xia}, \citenamefont
  {Wang},\ and\ \citenamefont {Liu}}]{song2017skyrmion}%
  \BibitemOpen
  \bibfield  {author} {\bibinfo {author} {\bibfnamefont {C.}~\bibnamefont
  {Song}}, \bibinfo {author} {\bibfnamefont {C.}~\bibnamefont {Jin}}, \bibinfo
  {author} {\bibfnamefont {J.}~\bibnamefont {Wang}}, \bibinfo {author}
  {\bibfnamefont {H.}~\bibnamefont {Xia}}, \bibinfo {author} {\bibfnamefont
  {J.}~\bibnamefont {Wang}},\ and\ \bibinfo {author} {\bibfnamefont
  {Q.}~\bibnamefont {Liu}},\ }\bibfield  {title} {\bibinfo {title}
  {Skyrmion-based multi-channel racetrack},\ }\href
  {https://aip.scitation.org/doi/10.1063/1.4994093} {\bibfield  {journal}
  {\bibinfo  {journal} {Appl. Phys. Lett.}\ }\textbf {\bibinfo {volume}
  {111}},\ \bibinfo {pages} {192413} (\bibinfo {year} {2017})}\BibitemShut
  {NoStop}%
\bibitem [{\citenamefont {Hoffmann}\ \emph {et~al.}(2017)\citenamefont
  {Hoffmann}, \citenamefont {Zimmermann}, \citenamefont {M{\"{u}}ller},
  \citenamefont {Sch{\"{u}}rhoff}, \citenamefont {Kiselev}, \citenamefont
  {Melcher},\ and\ \citenamefont {Bl{\"{u}}gel}}]{Hoffmann2017}%
  \BibitemOpen
  \bibfield  {author} {\bibinfo {author} {\bibfnamefont {M.}~\bibnamefont
  {Hoffmann}}, \bibinfo {author} {\bibfnamefont {B.}~\bibnamefont
  {Zimmermann}}, \bibinfo {author} {\bibfnamefont {G.~P.}\ \bibnamefont
  {M{\"{u}}ller}}, \bibinfo {author} {\bibfnamefont {D.}~\bibnamefont
  {Sch{\"{u}}rhoff}}, \bibinfo {author} {\bibfnamefont {N.~S.}\ \bibnamefont
  {Kiselev}}, \bibinfo {author} {\bibfnamefont {C.}~\bibnamefont {Melcher}},\
  and\ \bibinfo {author} {\bibfnamefont {S.}~\bibnamefont {Bl{\"{u}}gel}},\
  }\bibfield  {title} {\bibinfo {title} {{Antiskyrmions stabilized at
  interfaces by anisotropic Dzyaloshinskii-Moriya interactions}},\ }\href
  {https://doi.org/10.1038/s41467-017-00313-0} {\bibfield  {journal} {\bibinfo
  {journal} {Nat. Commun.}\ }\textbf {\bibinfo {volume} {8}},\ \bibinfo {pages}
  {1} (\bibinfo {year} {2017})}\BibitemShut {NoStop}%
\bibitem [{\citenamefont {Jena}\ \emph {et~al.}(2020)\citenamefont {Jena},
  \citenamefont {G{\"o}bel}, \citenamefont {Ma}, \citenamefont {Kumar},
  \citenamefont {Saha}, \citenamefont {Mertig}, \citenamefont {Felser},\ and\
  \citenamefont {Parkin}}]{Jena2020}%
  \BibitemOpen
  \bibfield  {author} {\bibinfo {author} {\bibfnamefont {J.}~\bibnamefont
  {Jena}}, \bibinfo {author} {\bibfnamefont {B.}~\bibnamefont {G{\"o}bel}},
  \bibinfo {author} {\bibfnamefont {T.}~\bibnamefont {Ma}}, \bibinfo {author}
  {\bibfnamefont {V.}~\bibnamefont {Kumar}}, \bibinfo {author} {\bibfnamefont
  {R.}~\bibnamefont {Saha}}, \bibinfo {author} {\bibfnamefont {I.}~\bibnamefont
  {Mertig}}, \bibinfo {author} {\bibfnamefont {C.}~\bibnamefont {Felser}},\
  and\ \bibinfo {author} {\bibfnamefont {S.~S.~P.}\ \bibnamefont {Parkin}},\
  }\bibfield  {title} {\bibinfo {title} {Elliptical bloch skyrmion chiral twins
  in an antiskyrmion system},\ }\href
  {https://doi.org/10.1038/s41467-020-14925-6} {\bibfield  {journal} {\bibinfo
  {journal} {Nat. Commun.}\ }\textbf {\bibinfo {volume} {11}},\ \bibinfo
  {pages} {1115} (\bibinfo {year} {2020})}\BibitemShut {NoStop}%
\bibitem [{\citenamefont {Zhang}\ \emph {et~al.}(2016)\citenamefont {Zhang},
  \citenamefont {Petford-Long},\ and\ \citenamefont
  {Phatak}}]{zhang2016creation}%
  \BibitemOpen
  \bibfield  {author} {\bibinfo {author} {\bibfnamefont {S.}~\bibnamefont
  {Zhang}}, \bibinfo {author} {\bibfnamefont {A.}~\bibnamefont
  {Petford-Long}},\ and\ \bibinfo {author} {\bibfnamefont {C.}~\bibnamefont
  {Phatak}},\ }\bibfield  {title} {\bibinfo {title} {Creation of artificial
  skyrmions and antiskyrmions by anisotropy engineering},\ }\href
  {https://www.nature.com/articles/srep31248} {\bibfield  {journal} {\bibinfo
  {journal} {Sci. Rep.}\ }\textbf {\bibinfo {volume} {6}},\ \bibinfo {pages}
  {1} (\bibinfo {year} {2016})}\BibitemShut {NoStop}%
\bibitem [{\citenamefont {Romming}\ \emph {et~al.}(2015)\citenamefont
  {Romming}, \citenamefont {Kubetzka}, \citenamefont {Hanneken}, \citenamefont
  {von Bergmann},\ and\ \citenamefont {Wiesendanger}}]{Romming2015}%
  \BibitemOpen
  \bibfield  {author} {\bibinfo {author} {\bibfnamefont {N.}~\bibnamefont
  {Romming}}, \bibinfo {author} {\bibfnamefont {A.}~\bibnamefont {Kubetzka}},
  \bibinfo {author} {\bibfnamefont {C.}~\bibnamefont {Hanneken}}, \bibinfo
  {author} {\bibfnamefont {K.}~\bibnamefont {von Bergmann}},\ and\ \bibinfo
  {author} {\bibfnamefont {R.}~\bibnamefont {Wiesendanger}},\ }\bibfield
  {title} {\bibinfo {title} {Field-dependent size and shape of single magnetic
  skyrmions},\ }\href {https://doi.org/10.1103/PhysRevLett.114.177203}
  {\bibfield  {journal} {\bibinfo  {journal} {Phys. Rev. Lett.}\ }\textbf
  {\bibinfo {volume} {114}},\ \bibinfo {pages} {177203} (\bibinfo {year}
  {2015})}\BibitemShut {NoStop}%
\bibitem [{\citenamefont {Tatara}\ \emph {et~al.}(2008)\citenamefont {Tatara},
  \citenamefont {Kohno},\ and\ \citenamefont {Shibata}}]{Tatara2008}%
  \BibitemOpen
  \bibfield  {author} {\bibinfo {author} {\bibfnamefont {G.}~\bibnamefont
  {Tatara}}, \bibinfo {author} {\bibfnamefont {H.}~\bibnamefont {Kohno}},\ and\
  \bibinfo {author} {\bibfnamefont {J.}~\bibnamefont {Shibata}},\ }\bibfield
  {title} {\bibinfo {title} {Microscopic approach to current-driven domain wall
  dynamics},\ }\href
  {https://doi.org/https://doi.org/10.1016/j.physrep.2008.07.003} {\bibfield
  {journal} {\bibinfo  {journal} {Phys. Rep.}\ }\textbf {\bibinfo {volume}
  {468}},\ \bibinfo {pages} {213} (\bibinfo {year} {2008})}\BibitemShut
  {NoStop}%
\bibitem [{\citenamefont {Bazaliy}\ \emph {et~al.}(1998)\citenamefont
  {Bazaliy}, \citenamefont {Jones},\ and\ \citenamefont {Zhang}}]{Bazaliy1998}%
  \BibitemOpen
  \bibfield  {author} {\bibinfo {author} {\bibfnamefont {Y.~B.}\ \bibnamefont
  {Bazaliy}}, \bibinfo {author} {\bibfnamefont {B.}~\bibnamefont {Jones}},\
  and\ \bibinfo {author} {\bibfnamefont {S.-C.}\ \bibnamefont {Zhang}},\
  }\bibfield  {title} {\bibinfo {title} {Modification of the landau-lifshitz
  equation in the presence of a spin-polarized current in colossal-and
  giant-magnetoresistive materials},\ }\href
  {https://journals.aps.org/prb/abstract/10.1103/PhysRevB.57.R3213} {\bibfield
  {journal} {\bibinfo  {journal} {Phys. Rev. B}\ }\textbf {\bibinfo {volume}
  {57}},\ \bibinfo {pages} {R3213} (\bibinfo {year} {1998})}\BibitemShut
  {NoStop}%
\bibitem [{\citenamefont {Zhang}\ and\ \citenamefont {Li}(2004)}]{Zhang2004}%
  \BibitemOpen
  \bibfield  {author} {\bibinfo {author} {\bibfnamefont {S.}~\bibnamefont
  {Zhang}}\ and\ \bibinfo {author} {\bibfnamefont {Z.}~\bibnamefont {Li}},\
  }\bibfield  {title} {\bibinfo {title} {Roles of nonequilibrium conduction
  electrons on the magnetization dynamics of ferromagnets},\ }\href
  {https://journals.aps.org/prl/abstract/10.1103/PhysRevLett.93.127204}
  {\bibfield  {journal} {\bibinfo  {journal} {Phys. Rev. Lett.}\ }\textbf
  {\bibinfo {volume} {93}},\ \bibinfo {pages} {127204} (\bibinfo {year}
  {2004})}\BibitemShut {NoStop}%
\bibitem [{\citenamefont {Lakshmanan}(2011)}]{Lakshmanan2011}%
  \BibitemOpen
  \bibfield  {author} {\bibinfo {author} {\bibfnamefont {M.}~\bibnamefont
  {Lakshmanan}},\ }\bibfield  {title} {\bibinfo {title} {The fascinating world
  of the landau-lifshitz-gilbert equation: an overview},\ }\href
  {https://doi.org/10.1098/rsta.2010.0319} {\bibfield  {journal} {\bibinfo
  {journal} {Philos. Trans. R. Soc. Lond. A}\ }\textbf {\bibinfo {volume}
  {369}},\ \bibinfo {pages} {1280} (\bibinfo {year} {2011})}\BibitemShut
  {NoStop}%
\bibitem [{\citenamefont {Huang}\ \emph {et~al.}(2017)\citenamefont {Huang},
  \citenamefont {Zhou}, \citenamefont {Chen}, \citenamefont {Shen},
  \citenamefont {Schmid}, \citenamefont {Liu},\ and\ \citenamefont
  {Wu}}]{Huang2017}%
  \BibitemOpen
  \bibfield  {author} {\bibinfo {author} {\bibfnamefont {S.}~\bibnamefont
  {Huang}}, \bibinfo {author} {\bibfnamefont {C.}~\bibnamefont {Zhou}},
  \bibinfo {author} {\bibfnamefont {G.}~\bibnamefont {Chen}}, \bibinfo {author}
  {\bibfnamefont {H.}~\bibnamefont {Shen}}, \bibinfo {author} {\bibfnamefont
  {A.~K.}\ \bibnamefont {Schmid}}, \bibinfo {author} {\bibfnamefont
  {K.}~\bibnamefont {Liu}},\ and\ \bibinfo {author} {\bibfnamefont
  {Y.}~\bibnamefont {Wu}},\ }\bibfield  {title} {\bibinfo {title}
  {{Stabilization and current-induced motion of antiskyrmion in the presence of
  anisotropic Dzyaloshinskii-Moriya interaction}},\ }\href
  {https://doi.org/10.1103/PhysRevB.96.144412} {\bibfield  {journal} {\bibinfo
  {journal} {Phys. Rev. B}\ }\textbf {\bibinfo {volume} {96}},\ \bibinfo
  {pages} {144412} (\bibinfo {year} {2017})}\BibitemShut {NoStop}%
\bibitem [{\citenamefont {Hoffmann}\ \emph {et~al.}(2021)\citenamefont
  {Hoffmann}, \citenamefont {Müller}, \citenamefont {Melcher},\ and\
  \citenamefont {Blügel}}]{Hoffmann2021}%
  \BibitemOpen
  \bibfield  {author} {\bibinfo {author} {\bibfnamefont {M.}~\bibnamefont
  {Hoffmann}}, \bibinfo {author} {\bibfnamefont {G.~P.}\ \bibnamefont
  {Müller}}, \bibinfo {author} {\bibfnamefont {C.}~\bibnamefont {Melcher}},\
  and\ \bibinfo {author} {\bibfnamefont {S.}~\bibnamefont {Blügel}},\
  }\bibfield  {title} {\bibinfo {title} {Skyrmion-antiskyrmion racetrack memory
  in rank-one dmi materials},\ }\href
  {https://www.frontiersin.org/article/10.3389/fphy.2021.769873} {\bibfield
  {journal} {\bibinfo  {journal} {Front. Phys.}\ }\textbf {\bibinfo {volume}
  {9}} (\bibinfo {year} {2021})}\BibitemShut {NoStop}%
\bibitem [{\citenamefont {Shimizu}\ \emph {et~al.}(2020)\citenamefont
  {Shimizu}, \citenamefont {Nagase}, \citenamefont {So}, \citenamefont
  {Kuwahara}, \citenamefont {Ikarashi},\ and\ \citenamefont
  {Nagao}}]{Shimizu2020}%
  \BibitemOpen
  \bibfield  {author} {\bibinfo {author} {\bibfnamefont {D.}~\bibnamefont
  {Shimizu}}, \bibinfo {author} {\bibfnamefont {T.}~\bibnamefont {Nagase}},
  \bibinfo {author} {\bibfnamefont {Y.-G.}\ \bibnamefont {So}}, \bibinfo
  {author} {\bibfnamefont {M.}~\bibnamefont {Kuwahara}}, \bibinfo {author}
  {\bibfnamefont {N.}~\bibnamefont {Ikarashi}},\ and\ \bibinfo {author}
  {\bibfnamefont {M.}~\bibnamefont {Nagao}},\ }\bibfield  {title} {\bibinfo
  {title} {Interaction between skyrmions and antiskyrmions in a coexisting
  phase of a heusler material},\ }\href
  {https://ui.adsabs.harvard.edu/abs/2020arXiv200807272S} {\ ,\ \bibinfo {eid}
  {arXiv:2008.07272} (\bibinfo {year} {2020})}\BibitemShut {NoStop}%
\bibitem [{\citenamefont {Fert}\ \emph
  {et~al.}(2013{\natexlab{b}})\citenamefont {Fert}, \citenamefont {Cros},\ and\
  \citenamefont {Sampaio}}]{Fert2013}%
  \BibitemOpen
  \bibfield  {author} {\bibinfo {author} {\bibfnamefont {A.}~\bibnamefont
  {Fert}}, \bibinfo {author} {\bibfnamefont {V.}~\bibnamefont {Cros}},\ and\
  \bibinfo {author} {\bibfnamefont {J.}~\bibnamefont {Sampaio}},\ }\bibfield
  {title} {\bibinfo {title} {Skyrmions on the track},\ }\href
  {https://doi.org/10.1038/nnano.2013.29} {\bibfield  {journal} {\bibinfo
  {journal} {Nat. Nanotechnol.}\ }\textbf {\bibinfo {volume} {8}},\ \bibinfo
  {pages} {152} (\bibinfo {year} {2013}{\natexlab{b}})}\BibitemShut {NoStop}%
\bibitem [{\citenamefont {Kumar}\ \emph {et~al.}(2020)\citenamefont {Kumar},
  \citenamefont {Kumar}, \citenamefont {Reehuis}, \citenamefont {Gayles},
  \citenamefont {Sukhanov}, \citenamefont {Hoser}, \citenamefont {Damay},
  \citenamefont {Shekhar}, \citenamefont {Adler},\ and\ \citenamefont
  {Felser}}]{Kumar2020}%
  \BibitemOpen
  \bibfield  {author} {\bibinfo {author} {\bibfnamefont {V.}~\bibnamefont
  {Kumar}}, \bibinfo {author} {\bibfnamefont {N.}~\bibnamefont {Kumar}},
  \bibinfo {author} {\bibfnamefont {M.}~\bibnamefont {Reehuis}}, \bibinfo
  {author} {\bibfnamefont {J.}~\bibnamefont {Gayles}}, \bibinfo {author}
  {\bibfnamefont {A.~S.}\ \bibnamefont {Sukhanov}}, \bibinfo {author}
  {\bibfnamefont {A.}~\bibnamefont {Hoser}}, \bibinfo {author} {\bibfnamefont
  {F.~m.~c.}\ \bibnamefont {Damay}}, \bibinfo {author} {\bibfnamefont
  {C.}~\bibnamefont {Shekhar}}, \bibinfo {author} {\bibfnamefont
  {P.}~\bibnamefont {Adler}},\ and\ \bibinfo {author} {\bibfnamefont
  {C.}~\bibnamefont {Felser}},\ }\bibfield  {title} {\bibinfo {title}
  {Detection of antiskyrmions by topological hall effect in heusler
  compounds},\ }\href {https://doi.org/10.1103/PhysRevB.101.014424} {\bibfield
  {journal} {\bibinfo  {journal} {Phys. Rev. B}\ }\textbf {\bibinfo {volume}
  {101}},\ \bibinfo {pages} {014424} (\bibinfo {year} {2020})}\BibitemShut
  {NoStop}%
\bibitem [{\citenamefont {Bouaziz}\ \emph {et~al.}(2021)\citenamefont
  {Bouaziz}, \citenamefont {Ishida}, \citenamefont {Lounis},\ and\
  \citenamefont {Bl\"ugel}}]{Bouaziz2021}%
  \BibitemOpen
  \bibfield  {author} {\bibinfo {author} {\bibfnamefont {J.}~\bibnamefont
  {Bouaziz}}, \bibinfo {author} {\bibfnamefont {H.}~\bibnamefont {Ishida}},
  \bibinfo {author} {\bibfnamefont {S.}~\bibnamefont {Lounis}},\ and\ \bibinfo
  {author} {\bibfnamefont {S.}~\bibnamefont {Bl\"ugel}},\ }\bibfield  {title}
  {\bibinfo {title} {Transverse transport in two-dimensional relativistic
  systems with nontrivial spin textures},\ }\href
  {https://doi.org/10.1103/PhysRevLett.126.147203} {\bibfield  {journal}
  {\bibinfo  {journal} {Phys. Rev. Lett.}\ }\textbf {\bibinfo {volume} {126}},\
  \bibinfo {pages} {147203} (\bibinfo {year} {2021})}\BibitemShut {NoStop}%
\bibitem [{\citenamefont {Tacchi}\ \emph {et~al.}(2017)\citenamefont {Tacchi},
  \citenamefont {Troncoso}, \citenamefont {Ahlberg}, \citenamefont {Gubbiotti},
  \citenamefont {Madami}, \citenamefont {{\AA}kerman},\ and\ \citenamefont
  {Landeros}}]{Tacchi2017}%
  \BibitemOpen
  \bibfield  {author} {\bibinfo {author} {\bibfnamefont {S.}~\bibnamefont
  {Tacchi}}, \bibinfo {author} {\bibfnamefont {R.~E.}\ \bibnamefont
  {Troncoso}}, \bibinfo {author} {\bibfnamefont {M.}~\bibnamefont {Ahlberg}},
  \bibinfo {author} {\bibfnamefont {G.}~\bibnamefont {Gubbiotti}}, \bibinfo
  {author} {\bibfnamefont {M.}~\bibnamefont {Madami}}, \bibinfo {author}
  {\bibfnamefont {J.}~\bibnamefont {{\AA}kerman}},\ and\ \bibinfo {author}
  {\bibfnamefont {P.}~\bibnamefont {Landeros}},\ }\bibfield  {title} {\bibinfo
  {title} {{Interfacial Dzyaloshinskii-Moriya Interaction in Pt/CoFeB Films:
  Effect of the Heavy-Metal Thickness}},\ }\href
  {https://doi.org/10.1103/PhysRevLett.118.147201} {\bibfield  {journal}
  {\bibinfo  {journal} {Phys. Rev. Lett.}\ }\textbf {\bibinfo {volume} {118}},\
  \bibinfo {pages} {147201} (\bibinfo {year} {2017})}\BibitemShut {NoStop}%
\bibitem [{\citenamefont {Cho}\ \emph {et~al.}(2015)\citenamefont {Cho},
  \citenamefont {Kim}, \citenamefont {Lee}, \citenamefont {Kim}, \citenamefont
  {Lavrijsen}, \citenamefont {Solignac}, \citenamefont {Yin}, \citenamefont
  {Han}, \citenamefont {{Van Hoof}}, \citenamefont {Swagten}, \citenamefont
  {Koopmans},\ and\ \citenamefont {You}}]{Cho2015}%
  \BibitemOpen
  \bibfield  {author} {\bibinfo {author} {\bibfnamefont {J.}~\bibnamefont
  {Cho}}, \bibinfo {author} {\bibfnamefont {N.~H.}\ \bibnamefont {Kim}},
  \bibinfo {author} {\bibfnamefont {S.}~\bibnamefont {Lee}}, \bibinfo {author}
  {\bibfnamefont {J.~S.}\ \bibnamefont {Kim}}, \bibinfo {author} {\bibfnamefont
  {R.}~\bibnamefont {Lavrijsen}}, \bibinfo {author} {\bibfnamefont
  {A.}~\bibnamefont {Solignac}}, \bibinfo {author} {\bibfnamefont
  {Y.}~\bibnamefont {Yin}}, \bibinfo {author} {\bibfnamefont {D.~S.}\
  \bibnamefont {Han}}, \bibinfo {author} {\bibfnamefont {N.~J.}\ \bibnamefont
  {{Van Hoof}}}, \bibinfo {author} {\bibfnamefont {H.~J.}\ \bibnamefont
  {Swagten}}, \bibinfo {author} {\bibfnamefont {B.}~\bibnamefont {Koopmans}},\
  and\ \bibinfo {author} {\bibfnamefont {C.~Y.}\ \bibnamefont {You}},\
  }\bibfield  {title} {\bibinfo {title} {{Thickness dependence of the
  interfacial Dzyaloshinskii-Moriya interaction in inversion symmetry broken
  systems}},\ }\href {https://doi.org/10.1038/ncomms8635} {\bibfield  {journal}
  {\bibinfo  {journal} {Nat. Commun.}\ }\textbf {\bibinfo {volume} {6}},\
  \bibinfo {pages} {1} (\bibinfo {year} {2015})}\BibitemShut {NoStop}%
\bibitem [{\citenamefont {Camosi}\ \emph {et~al.}(2017)\citenamefont {Camosi},
  \citenamefont {Rohart}, \citenamefont {Fruchart}, \citenamefont {Pizzini},
  \citenamefont {Belmeguenai}, \citenamefont {Roussign{\'{e}}}, \citenamefont
  {Stashkevich}, \citenamefont {Cherif}, \citenamefont {Ranno}, \citenamefont
  {{De Santis}},\ and\ \citenamefont {Vogel}}]{Camosi2017}%
  \BibitemOpen
  \bibfield  {author} {\bibinfo {author} {\bibfnamefont {L.}~\bibnamefont
  {Camosi}}, \bibinfo {author} {\bibfnamefont {S.}~\bibnamefont {Rohart}},
  \bibinfo {author} {\bibfnamefont {O.}~\bibnamefont {Fruchart}}, \bibinfo
  {author} {\bibfnamefont {S.}~\bibnamefont {Pizzini}}, \bibinfo {author}
  {\bibfnamefont {M.}~\bibnamefont {Belmeguenai}}, \bibinfo {author}
  {\bibfnamefont {Y.}~\bibnamefont {Roussign{\'{e}}}}, \bibinfo {author}
  {\bibfnamefont {A.}~\bibnamefont {Stashkevich}}, \bibinfo {author}
  {\bibfnamefont {S.~M.}\ \bibnamefont {Cherif}}, \bibinfo {author}
  {\bibfnamefont {L.}~\bibnamefont {Ranno}}, \bibinfo {author} {\bibfnamefont
  {M.}~\bibnamefont {{De Santis}}},\ and\ \bibinfo {author} {\bibfnamefont
  {J.}~\bibnamefont {Vogel}},\ }\bibfield  {title} {\bibinfo {title}
  {{Anisotropic Dzyaloshinskii-Moriya interaction in ultrathin epitaxial
  Au/Co/W(110)}},\ }\href {https://doi.org/10.1103/PhysRevB.95.214422}
  {\bibfield  {journal} {\bibinfo  {journal} {Phys. Rev. B}\ }\textbf {\bibinfo
  {volume} {95}},\ \bibinfo {pages} {214422} (\bibinfo {year}
  {2017})}\BibitemShut {NoStop}%
\bibitem [{\citenamefont {Litzius}\ \emph {et~al.}(2017)\citenamefont
  {Litzius}, \citenamefont {Lemesh}, \citenamefont {Krüger}, \citenamefont
  {Bassirian}, \citenamefont {Caretta}, \citenamefont {Richter}, \citenamefont
  {Büttner}, \citenamefont {Sato}, \citenamefont {Tretiakov}, \citenamefont
  {Förster}, \citenamefont {Reeve}, \citenamefont {Weigand}, \citenamefont
  {Bykova}, \citenamefont {Stoll}, \citenamefont {Schütz}, \citenamefont
  {Beach},\ and\ \citenamefont {Kläui}}]{Litzius2017}%
  \BibitemOpen
  \bibfield  {author} {\bibinfo {author} {\bibfnamefont {K.}~\bibnamefont
  {Litzius}}, \bibinfo {author} {\bibfnamefont {I.}~\bibnamefont {Lemesh}},
  \bibinfo {author} {\bibfnamefont {B.}~\bibnamefont {Krüger}}, \bibinfo
  {author} {\bibfnamefont {P.}~\bibnamefont {Bassirian}}, \bibinfo {author}
  {\bibfnamefont {L.}~\bibnamefont {Caretta}}, \bibinfo {author} {\bibfnamefont
  {K.}~\bibnamefont {Richter}}, \bibinfo {author} {\bibfnamefont
  {F.}~\bibnamefont {Büttner}}, \bibinfo {author} {\bibfnamefont
  {K.}~\bibnamefont {Sato}}, \bibinfo {author} {\bibfnamefont {O.~A.}\
  \bibnamefont {Tretiakov}}, \bibinfo {author} {\bibfnamefont {J.}~\bibnamefont
  {Förster}}, \bibinfo {author} {\bibfnamefont {R.~M.}\ \bibnamefont {Reeve}},
  \bibinfo {author} {\bibfnamefont {M.}~\bibnamefont {Weigand}}, \bibinfo
  {author} {\bibfnamefont {I.}~\bibnamefont {Bykova}}, \bibinfo {author}
  {\bibfnamefont {H.}~\bibnamefont {Stoll}}, \bibinfo {author} {\bibfnamefont
  {G.}~\bibnamefont {Schütz}}, \bibinfo {author} {\bibfnamefont {G.~S.~D.}\
  \bibnamefont {Beach}},\ and\ \bibinfo {author} {\bibfnamefont
  {M.}~\bibnamefont {Kläui}},\ }\bibfield  {title} {\bibinfo {title} {Skyrmion
  hall effect revealed by direct time-resolved x-ray microscopy},\ }\href
  {https://doi.org/10.1038/nphys4000} {\bibfield  {journal} {\bibinfo
  {journal} {Nat. Phys.}\ }\textbf {\bibinfo {volume} {13}},\ \bibinfo {pages}
  {170} (\bibinfo {year} {2017})}\BibitemShut {NoStop}%
\bibitem [{\citenamefont {Zhang}\ \emph {et~al.}(2015)\citenamefont {Zhang},
  \citenamefont {Zhao}, \citenamefont {Fangohr}, \citenamefont {Liu},
  \citenamefont {Xia}, \citenamefont {Xia},\ and\ \citenamefont
  {Morvan}}]{Zhang2015}%
  \BibitemOpen
  \bibfield  {author} {\bibinfo {author} {\bibfnamefont {X.}~\bibnamefont
  {Zhang}}, \bibinfo {author} {\bibfnamefont {G.~P.}\ \bibnamefont {Zhao}},
  \bibinfo {author} {\bibfnamefont {H.}~\bibnamefont {Fangohr}}, \bibinfo
  {author} {\bibfnamefont {J.~P.}\ \bibnamefont {Liu}}, \bibinfo {author}
  {\bibfnamefont {W.~X.}\ \bibnamefont {Xia}}, \bibinfo {author} {\bibfnamefont
  {J.}~\bibnamefont {Xia}},\ and\ \bibinfo {author} {\bibfnamefont {F.~J.}\
  \bibnamefont {Morvan}},\ }\bibfield  {title} {\bibinfo {title}
  {Skyrmion-skyrmion and skyrmion-edge repulsions in skyrmion-based racetrack
  memory},\ }\href {https://doi.org/10.1038/srep07643} {\bibfield  {journal}
  {\bibinfo  {journal} {Sci. Rep.}\ }\textbf {\bibinfo {volume} {5}},\ \bibinfo
  {pages} {7643} (\bibinfo {year} {2015})}\BibitemShut {NoStop}%
\end{thebibliography}
%

\appendix
\clearpage
\onecolumngrid
\beginsupplement
\section*{Appendix A: Influence of the skyrmion Hall effect}
\begin{figure*}[!htbp]\centering
	\includegraphics[width=0.6\linewidth, height=0.6\textheight,keepaspectratio]{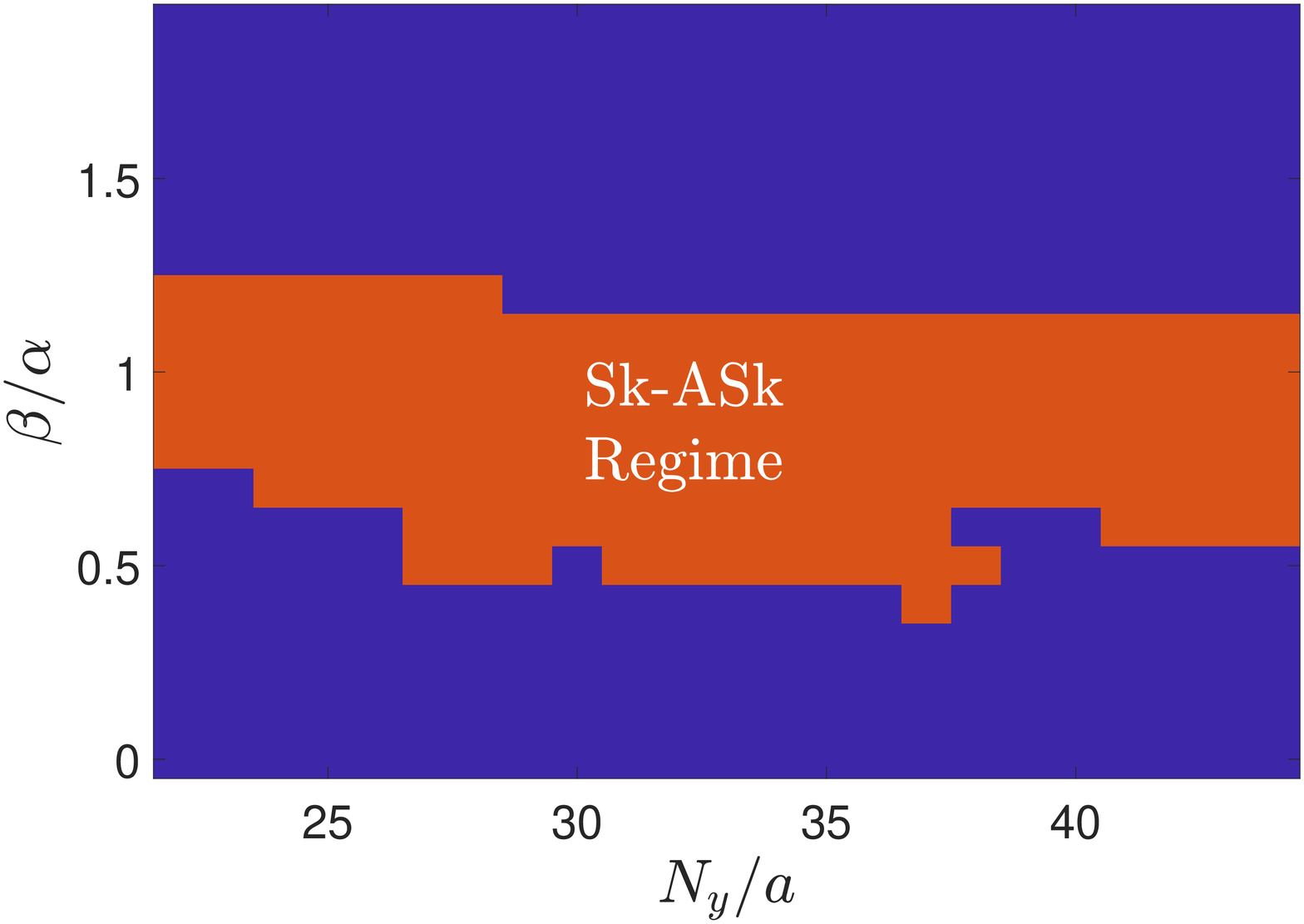}
	\caption{Stability diagram of an array of six SKs and ASKs (3 of each species) in dependence on the ratio of non-adiabaticity constant $\beta$ and damping constant $\alpha$ and the racetrack width $N_y$ for $\alpha=0.1$. If $\beta$ deviates too much from $\alpha$, the resulting skyrmion Hall effect leads to a destruction of the created quasiparticles at the edges during the current pulses.}
	\label{SUP:1}
\end{figure*}

Current driven (anti-)skyrmions experience the skyrmion Hall effect, see Refs. \cite{Nagaosa2013,jiang2017direct,Litzius2017}, which induces a motion perpendicular to the direction of the applied current. 
This sidewards motion is often undesired, because it prevents a movement of (anti-)skyrmions along a straight line and can, if too strong, even lead to a destruction of magnetic structures at the edge, see Refs. \cite{Zhang2015,iwasaki2013current}.
In the main text, we have considered $\beta=\alpha$, such that no skyrmion Hall effect occurs \cite{stier2017skyrmion}.
As this condition is not met by most of the materials, we study here the regime $\beta\neq\alpha$ to explore the robustness of the proposed concept. Figure \ref{SUP:1} depicts the stability regime of a SK-ASK array in dependence on $\beta$ and on the sample width $N_y$. With the  same interaction parameters considered in the main text and with taking $D_x=0$ and $\tau=150$~ps, a ratio of $0.8<\beta/\alpha<1.2$ allows for the creation of a stable array for nearly all sample widths.
If $\beta$ deviates too strongly from $\alpha$, the sidewards motion of (anti-)skyrmions leads to a destruction of the quasiparticles at the ferromagnetic boundary. When pushed against the boundary, the (anti-)skyrmions shrink below a critical radius and decay. During the periods where no current is applied ($v_s=0$), the (anti-)skyrmions regrow to their stable size.
A fine-tuning of $v_s$ and $\tau$ is expected to enlarge the regime of $\frac{\beta}{\alpha}$ that allows for a stable SK-ASK propulsion. Furthermore, a break during the current pulses, giving the (anti-)skyrmions time to recover their size, can increase the regime of stability. It is also possible to use advanced material engineering to avoid the skyrmion Hall effect at all \cite{plettenberg2020steering}.\\
%
\section*{Appendix B: Current-less skyrmion-antiskyrmion racetrack}
The skyrmion Hall effect can lead to limitations of the proposed SK-ASK racetrack concept.
In the following, we introduce two currentless setups of the SK-ASK data storage device which rely on the same basic elements, i.e., a (anti-)skyrmion creation by a rotation of the edge magnetization and a stabilization due to anisotropic DMI, but which work with an adapted creation protocol. The first concept relies on the successive creation of SKs and ASKs at the short edge of the racetrack. The second one realizes the simultaneous creation of localized SKs and ASKs at various creation areas.
\subsection*{B.1: Creation by Gaussian Rotation}
\begin{figure*}
	\centering
	\begin{tabular}{c c}
		\includegraphics[width=0.45\linewidth, height=\textheight,keepaspectratio]{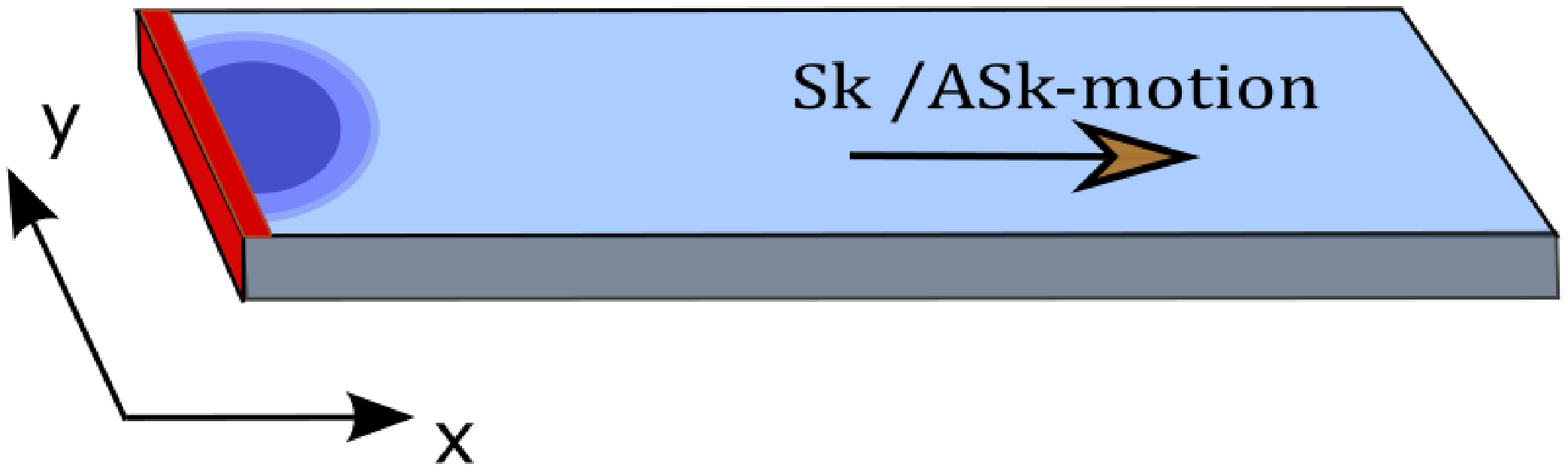} 
		\begin{picture}(0,0)
		\put(-200,120){{\large (a)}}
		\end{picture}
		\begin{picture}(0,0)
		\put(-0,120){{\large (b)}}
		\end{picture} &
		\includegraphics[width=0.45\linewidth,height=\textheight,keepaspectratio]{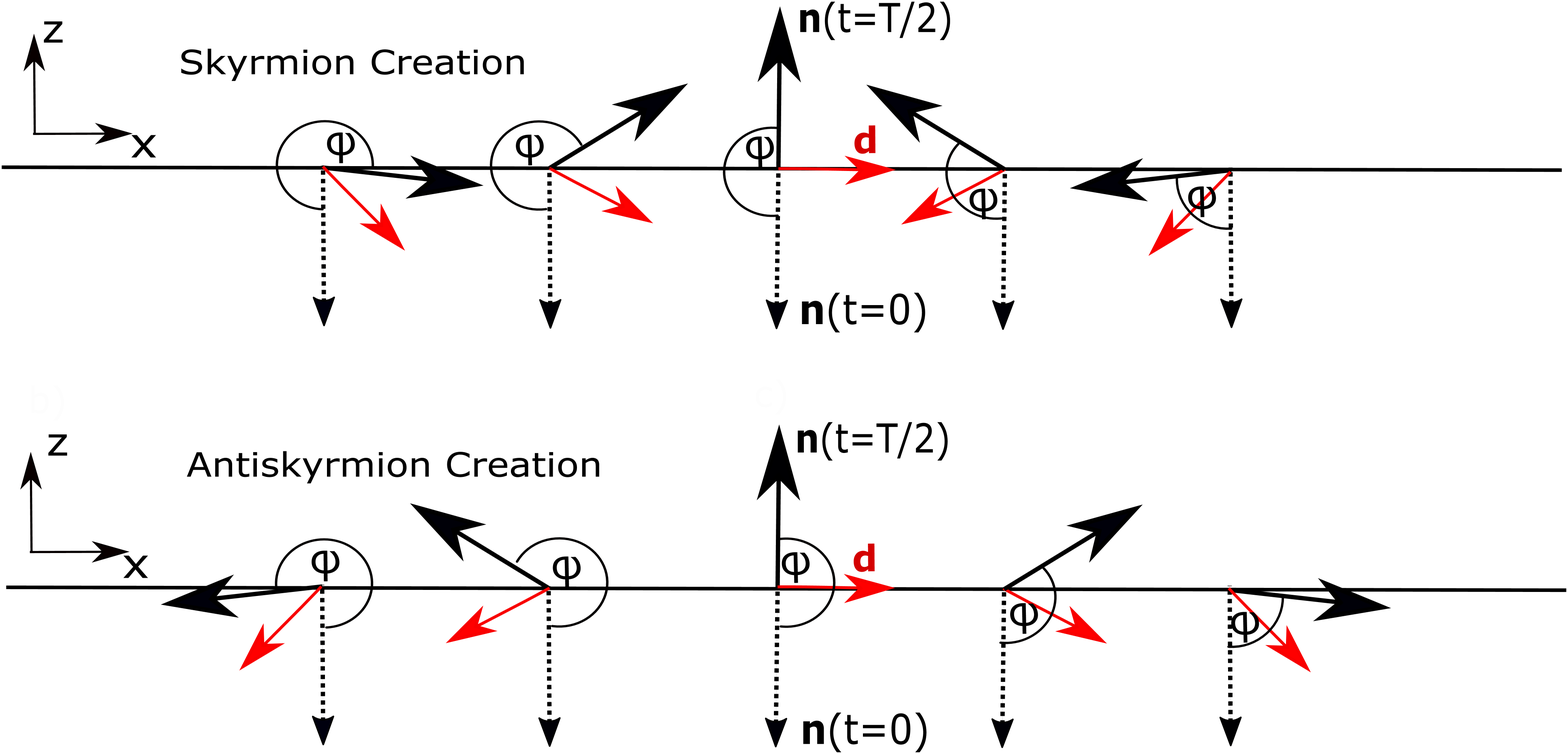}
		
	\end{tabular}
	\caption{(a) Schematic of currentless SK-ASK racetrack. The magnetization at the short edge (red region) is rotated according to Eq. (\ref{RotGauss}). A successive creation of quasiparticles leads to a shift of the already created quasiparticles along the racetrack due to the repulsive interactions along the $x$ direction. (b) Schematic of the edge rotation scheme to create a SK (upper row) and an ASK (lower row). The edge magnetization (black arrows) is rotated around the spatially dependent rotation axis (red arrows). Depicted are the magnetic moments $\textbf{n}$ in their initial configuration (dashed line) and after half a rotation (solid line) of period $T$.}
	\label{SUP:2}
\end{figure*}

By a more sophisticated rotation scheme it is possible to create SKs and ASKs at the short edge of the racetrack, as indicated in Fig. \ref{SUP:2} (a). This has the advantage that SKs and ASKs can, in analogy to the pure SK racetrack in Ref. \cite{Schaffer2020} push each other through the racetrack without a current due to their repulsive interactions along the $x$ direction.
To allow for the controlled creation of both magnetic species, the rotation scheme needs to induce a tilt  of the magnetic moments along the $y$ direction.
The used rotation scheme is an adapted Gauss rotation, discussed in the Supplemental Material of Ref. \cite{Schaffer2020}.
We take the inspiration for this rotation from the $z$-component of the profile of the magnetization of a SKs which can be approximately fitted by a Gauss function
\begin{equation}
m_z(y)=-2\exp\left[-\left(\frac{y-y_0}{\sigma}\right)^2\right]+1,
\end{equation}
where $y_0$ is the coordinate of the SK core, $y$ the position of the magnetic moments and $\sigma$ a measure for the width of the Gauss function.
The magnetization at the sample edge is rotated around a spatially dependent rotation axis such that the profile magnetization of SKs and ASKs is recovered after half a rotation, as sketched in Fig. \ref{SUP:2} (b). The angle $\phi$ between the initial  position of the magnetic moments and the position after half a rotation (Fig. \ref{SUP:2} (b)) is given by
\begin{equation} \label{RotGauss}
\cos\phi=-2\exp\left[-\left(\frac{y-y_0}{\sigma}\right)^2\right]+1.
\end{equation}
For the creation of an ASK, the $y-$component of the rotation axis is mirrored $n_y \to - n_y$ according to the rotation creating a SK. 
The rotation schemes then resemble the shift of a (anti-)skyrmion over the edge.
Thus, a realization of these rotations could be a coupling to a movable stable SK in a different neighboring layer. The rotation then is achieved by a movement of the ``seed SK'' over the edge of the racetrack part of the sample. This is analogous to a hard magnet writing its information to a soft magnet.

\subsection*{B.2: Creation by multiple rotation areas}
\begin{figure*}
	\centering
	\includegraphics[width=0.7\linewidth,keepaspectratio]{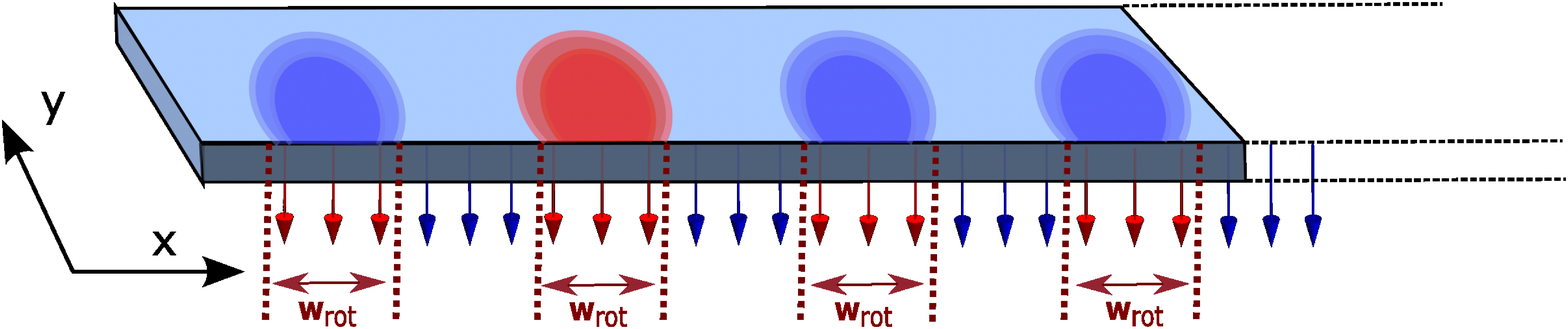}
	\caption{Second concept of a currentless SK-ASK data storage device. The creation of the magnetic structures takes place at various creation areas of width $\text{w}_{\rm{rot}}$ at the long edge of the sample. In between these creation areas, the boundary magnetic moments are fixed along the $-z$-direction (blue arrows). Depending on the rotation sense a SK (blue) or an ASK (red) is created.}
	\label{SUP:3}
\end{figure*}

A second approach of a currentless SK-ASK data storage device relies on multiple creation areas. The creation occurs at multiple positions at the long edge, as shown in Fig. \ref{SUP:3}. The edge magnetic moments are rotated by a uniform rotation, introduced in the main text.
In this concept, SKs and ASKs are not supposed to move along the racetrack but to stay positioned close to the creation and deletion area. This would request a simultaneous initialization of the whole sample to avoid drifts of the created quasiparticles. Then, the repulsive interactions between SKs and ASKs along the $x$ direction fixes the position of each magnetic quasiparticle. Systematic deletion and creation processes can alter the written data.
In contrast to the proposed racetrack concepts, this approach does not rely on mobile data carriers during the writing process. Of course, the data carriers could still be moved by a current to the final read out area.
\end{document}